\documentclass[%
 reprint,
 amsmath,amssymb,
 aps,
floatfix,
]{revtex4-2}

\usepackage{graphicx}
\usepackage{dcolumn}
\usepackage{bm}
\usepackage{natbib}
\usepackage{xcolor}

\usepackage{hyperref}

\begin{document}

\preprint{APS/123-QED}

\title{Constraining parameters of coalescing stellar mass binary black hole systems with the Einstein Telescope alone}

\author{Neha Singh}
\email{nsingh@astrouw.edu.pl}
\affiliation{Astronomical Observatory, University of Warsaw,\\ Al. Ujazdowskie 4, 00-478 Warsaw, Poland}
\author{Tomasz Bulik}
 \email{tb@astrouw.edu.pl}
\affiliation{Astronomical Observatory, University of Warsaw,\\ Al. Ujazdowskie 4, 00-478 Warsaw, Poland}

\date{\today}

\begin{abstract}

The Einstein Telescope (ET) is the future third generation  gravitational wave detector consisting of three independent interferometers arranged in a triangular configuration, with the sensitivity large enough to be able to detect stellar mass black holes even beyond the redshift of 10.
In this paper, we analyze the capabilities of ET as a standalone instrument and not part of a network.
We show that the analysis of detection of binary coalescences in the three individual interferometers of the ET allow us to weakly constrain the sky location of the source.  We present the analysis that leads to the constraints on the redshift and source frame chirp mass of detected binaries. We show that these values can be estimated with the accuracy comparable to the one expected from networks of gravitational wave interferometers. Thus, we show that the ET as a single instrument is able to break the redshift - chirp mass degeneracy and is therefore a valuable tool to explore properties of populations of merging compact object binaries. 
\end{abstract}


\maketitle

\section{Introduction} \label{sec:intro}

Gravitational wave astronomy is in its nascent state; however, it is growing very fast.  The first detection of a binary black hole (BBH) coalescence \cite{2016PhRvL.116f1102A} has opened the field for more information about the Universe.  The detection of a binary neutron star along with a GRB \cite{BNSGRB} has opened a path to multimessenger studies of compact object coalescences. 
At the moment of writing, there are already numerous gravitational wave binary detections  \cite{GWTC-2, GWTC-1, GW190412, GW190521, BNSGRB, GW190425, GW190814}. However, we expect that these numbers will significantly increase after the LIGO VIRGO O3b run results are published. The next generation instruments like the Einstein Telescope (ET) \cite{Hild2011, 2010CQGra..27s4002P}  or Cosmic Explorer (CE) \cite{PhysRevD.91.082001, Abbott2017, 2019BAAS...51g..35R}  will allow us to detect stellar binary black hole (BBH) coalescences almost in the entire Universe and binary neutron stars up to the redshift of $z\approx 1$.  Given the current estimate of the BBH rate density, we expect that the ET will detect a BBH every minute \cite{2019ApJ...882L..24A}. The population studies of these objects will require an estimate of masses and distances. Therefore, it is important to develop algorithms that will allow us to make such estimates. The issue has already been considered by  \citet{2018PhRvD..97f4031Z} and \citet{2018PhRvD..97l3014C}  who used the Fisher information matrix to study the effects of the time-dependent detector response due to the Earth’s rotation on long duration signals from systems such as binary neutron star (BNS) and neutron star - black hole (NSBH) for estimating the uncertainties in the measurement of signal parameters using 3G detectors such as the ET and CE. \citet{PhysRevD.95.064052} presented a detailed study of short duration signals from binary black holes using the Einstein Telescope as part of a network of instruments. Recently,  \citet{2010CQGra..27s4002P} described the capability of the ET as a single instrument and as part of a network, to analyze both long and short duration signals for equal-mass, non-spinning binaries using Fisher-matrix analysis. In this paper, we analyze the case when ET acts as a single instrument to detect short signals, i.e. neglecting the rotation of the Earth.

One of the fundamental questions that the ET will be able to address relates to the populations of compact object binaries. From the point of view of measuring the properties of population, we would like to know the distances, or redshifts and masses of the events. Since the population is at cosmological distances, the objects will be distributed uniformly on the sky. Population studies do not require exact positions on the sky or polarizations of individual binaries; however, quite a wealth of information can be inferred from the distributions of measured parameters like the chirp mass \cite{2004A&A...415..407B}. While accurate positions may be needed for counterpart searches, we do not think that it is realistic that such searches will be undertaken for every binary ET detects. Therefore, for the purpose of studying the properties of merging compact object populations, one needs the estimate of distances or redshifts, and source frame chirp masses, while the estimates of the sky location is not relevant. 
Moreover, one can perform efficient population studies even if redshifts and source frame chirp masses have uncertainties up to 20\%-30\%, provided that we have a large statistics of sources.
In this paper, we present a method to estimate redshifts and source frame chirp masses using the ET as a standalone instrument. 

We assume that the ET  will have its  design sensitivity \cite{Hild2011} and investigate the prospects of using the ET as a single instrument. In particular, we  analyze the capabilities of the ET to constrain distances to BBH coalescences and their properties such as the chirp mass and redshift or luminosity distance. While triangulation cannot be used to localize the source for a detection of a binary coalescence by a single interferometer, the antenna pattern of three detectors can always be used to roughly constrain the location of the source. It has been already noticed by \citet{LeeFinn93} that a single interferometer detection allows us to constrain the distance using the distribution of probability of the detector-reach as a function of the position in the sky and polarization of the gravitational wave. In the case of the ET working as a single instrument, there will be three colocated interferometers, with different orientations providing three antenna patterns for the  analysis. In this work, we study the effectiveness of  detection of a binary coalescence with three interferometers in the ET  to constrain its luminosity distance, redshift and the chirp mass. We also discuss the possible constraints on individual masses.

\section{Detector and signal characteristics }

\subsection{Einstein telescope}
We use the currently planned  design of the ET-D \cite{Hild2011}, consisting of three overlapping interferometers, arranged in an equilateral configuration with arm-opening angles of $60^{\circ}$, with the ET-D noise curve \cite{Hild2011}. The response function of each detector in the ET  $F_{(+,\times)i}$, $i=\{1,2,3\}$ for the two GW polarizations are

\begin{subequations}\label{antenna}

\begin{equation}
\begin{split}
    F_{+1}\left(\theta,\phi,\psi\right) = \frac{\sqrt{3}}{4}\left(1+\cos^{2}\left(\theta\right)\right)\cos\left(2\phi\right)\cos\left(2\psi\right) \\
     -\frac{\sqrt{3}}{2}\cos\left(\theta\right)\sin\left(2\phi\right)\sin\left(2\psi\right),
\end{split}
\end{equation}

\begin{equation}
\begin{split}
    F_{\times1}\left(\theta,\phi,\psi\right) = \frac{\sqrt{3}}{4}\left(1+\cos^{2}\left(\theta\right)\right)\cos\left(2\phi\right)\sin\left(2\psi\right)\\
    + \frac{\sqrt{3}}{2}\cos\left(\theta\right)\sin\left(2\phi\right)\cos\left(2\psi\right),
\end{split}
\end{equation}

\begin{equation}
    F_{(+,\times)2}\left(\theta,\phi,\psi\right)=F_{(+,\times)1}\left(\theta, \phi + 2\pi /3, \psi\right),
\end{equation}

\begin{equation}
    F_{(+,\times)3}\left(\theta,\phi,\psi\right)=F_{(+,\times)1}\left(\theta, \phi + 4\pi /3, \psi\right),
\end{equation}
\end{subequations}
where $\theta,\phi$ denote the sky position and $\psi$ is the polarization angle.
Dropping the subscript that denotes the individual interferometer for convenience, $F_+, F_\times$ being the antenna response function of one of the three detectors in ET, the strain on the detector is given as

\begin{equation}\label{h_t_ant}
    h(t)= F_+h_+(t+t_c-t_0) + F_{\times}h_\times(t+t_c-t_0)
\end{equation}
where $t_0$ is the time of coalescence in the detector frame, and $t_0 -t_c$ is the travel time from the source to the detector.

\subsection{Waveform}
In the following analysis, we consider the ET alone to detect the gravitational radiation from an inspiraling BBH system. Such systems are also known as chirping binaries since the two polarizations of the gravitational wave produced by the gravitational-wave signal from a compact binary inspiral have a monotonically increasing frequency and amplitude as the orbital motion radiates gravitational wave energy. As described in Sec 3 of \cite{Findchirp}, the two polarizations for $t< t_c$  of the waveform for a binary with the chirp mass $\mathcal{M}$, merging at a distance $D_{L}$ are given as

\begin{subequations}\label{hpluscross_ant}

\begin{equation}
\begin{split}
    h_{+}(t) = -\frac{1+\cos^{2}\iota}{2}\left(\frac{G \mathcal{M} }{c^2D_{L}}\right)\left(\frac{t_c-t}{G\mathcal{M} /c^3}\right)^{-1/4}\\
    \times \cos\left[2\Phi_c + 2\Phi \left(t-t_c ; M,\mu\right)\right],
\end{split}
\end{equation}
\begin{equation}
\begin{split}
    h_{\times}(t) = -\cos\iota \left(\frac{G \mathcal{M}}{c^2D_{L}}\right)\left(\frac{t_c-t}{G\mathcal{M}/c^3}\right)^{-1/4}\\
    \times \sin\left[2\Phi_c + 2\Phi \left(t-t_c ; M,\mu\right)\right],
\end{split}
\end{equation}
\end{subequations}
where, $\iota$ is the angle between the direction to the observer and the orbital angular momentum axis of the binary system, and $\mu$ is the reduced mass of the binary system. $\Phi \left(t-t_c ; M,\mu\right)$ is orbital phase of the binary. For a binary system composed of component masses $m_1$ and $m_2$, $\mathcal{M} = (m_1m_2)^{3/5}/M^{1/5}$, where $M = m_1+m_2$. As defined in \cite{Findchirp}, $t_{c}$ and $\Phi_{c}$ denote the time and phase of the termination of the waveform. 
Substituting Eq. (\ref{hpluscross_ant}) in (\ref{h_t_ant}), we get

\begin{equation}\label{ht}
\begin{split}
    h(t)= -\left(\frac{G\mathcal{M}}{c^2}\right)\left(\frac{\Theta}{4D_L}\right)\left(\frac{t_0-t}{5G\mathcal{M}/c^3}\right)^{-1/4}\\
    \times \cos\left[2\Phi_0 + 2\Phi \left(t-t_c ; M,\mu\right)\right]
\end{split}
\end{equation}
where 
\begin{equation}\label{theta}
    \Theta\equiv 2 \left[F_{+}^{2}\left(1+\cos^{2}\iota\right)^{2} + 4F_{\times}^{2}\cos^{2}\iota\right]^{1/2}
\end{equation}
with $0<\Theta<4$ , and $\iota$ is the angle of inclination and as in \cite{Findchirp},

\begin{equation}\label{snr_phase}
    2\Phi_0 = 2\Phi_c - \arctan\left(\frac{2F_\times\cos\iota}{F_+\left(1\cos^2\iota\right)}\right)
\end{equation}
The Fourier transform of the signal amplitude of GWs from an inspiraling binary system assuming that the time of the signal in the detector bandwidth is short enough to ignore the change in the orientation of the detector due to rotation of the Earth, can be written as \cite{PhysRevD.44.3819, TaylorGair2012, 2010ApJ...716..615O}
\begin{equation}
    |\tilde{h}(f)|= \frac{2c}{D_L}\left(\frac{5G\mu}{96c^3}\right)^{1/2}\left(\frac{GM}{\pi^2c^3}\right)^{1/3}\left(\frac{\Theta}{4}\right)f^{-7/6}\label{h_fourier}
\end{equation}
The match-filtering signal to noise ratio $\rho$ for each of the three ET detector assuming that they have identical noise will be \cite{TaylorGair2012, LeeFinn96}:
\begin{equation}\label{snr}
\rho_i \approx 8 \Theta_i \frac{r_{0_i}}{D_{L}}
\left(\frac{\mathcal{M}_{z}}{\mathcal{M}_{BNS}}\right)^{5/6}\sqrt{\zeta\left(f_{max}\right)} 
\end{equation}
where  $\mathcal{M}_{z}= (1+z)\mathcal{M} $ is the redshifted chirp mass and $\mathcal{M}_{BNS}\approx 1.218 M_\odot $  is the chirp mass of a equal mass binary with each component of mass $1.4 M_{\odot}$,
\begin{equation}\label{zetafunc}
\zeta\left(f_{max}\right) = \frac{1}{x_{7/3}}\int^{2f_{max}}_{1}\frac{df \left( \pi M_{\odot}\right)^{2}}{\left(\pi f M_{\odot}\right)^{7/3}S_{h}\left(f\right)}
\end{equation}
$S_{h}\left(f\right)$ is the power spectral density (PSD) for ET-D configuration,  and $x_{7/3}$ is

\begin{equation}\label{x_7_3}
x_{7/3} = \int^{\infty}_{1}\frac{df \left( \pi M_{\odot}\right)^{2}}{\left(\pi f M_{\odot}\right)^{7/3}S_{h}\left(f\right)}.
\end{equation}
The characteristic distance sensitivity $r_0$ is:

\begin{equation}\label{detreach}
r^{2}_{0} = \frac{5}{192 \pi}\left(\frac{3 G}{20}\right)^{5/3}x_{7/3}\frac{M^{2}_{\odot}}{c^{3}},
\end{equation}
and $f_{max}$ is the frequency at the end of the inspiral,

\begin{equation}\label{fmax}
f_{max} = 785\left(\frac{M_{BNS}}{M(1+z)}\right)  \;\rm Hz
\end{equation}
where $M_{BNS}=2.8 M_\odot$ is the total mass of an equal mass binary with each component of mass $1.4 M_{\odot}$. Given the combined signal from three detectors, we can define the combined effective signal to noise ratio,

\begin{equation}\label{snreff}
\rho_{eff} = 8 \Theta_{eff} \frac{r_{0_i}}{D_{L}}\left(\frac{\mathcal{M}_{z}}{1.2 M_{\odot}}\right)^{5/6}\sqrt{\zeta\left(f_{max}\right)} 
\end{equation}
where the effective antenna response function  $\Theta_{eff}$ is

\begin{equation}\label{thetaeff}
\Theta_{eff} = \left(\Theta_{1}^{2} + \Theta_{2}^{2} + \Theta_{3}^{2}\right)^{1/2}.
\end{equation}

\section{The plan of the analysis} \label{sec:plan}

We assume that in each  detection of merger, the observables $D$ for a coalescing binary systems are  
(a) The three signal to noise ratios (SNRs) $\rho_{i}$ defined by Eq.(\ref{snr}). (b) The phase of the strain $\Phi_{o,i}$ for $i = (1,2,3)$ corresponding to three detectors, as defined in Eq.(\ref{snr_phase}). The quantity $\Phi_0$ is the best match phase obtained by maximizing the matched - filter output over the phase of the strain $h(t)$. The details are given in \cite{Findchirp}. (c) The redshifted chirp mass $\mathcal{M}_{z}$. (d) The frequency at the end of the inspiral, taken to correspond to the innermost stable circular orbit, $f_{max}$. We will use the signal to noise ratios in each interferometer to constrain the value of the effective antennae pattern. In order to do this, we analyze the value of the ratios of the SNRs:
  $\rho_{21} \equiv \rho_2/\rho_1$ and $\rho_{31} \equiv \rho_3/\rho_1$.
We note that using Eq. (\ref{snr}) these ratios are
\begin{equation}\label{theta-ratios}
     \rho_{21} = \frac{\Theta_2}{\Theta_1}\equiv\Theta_{21} \ \ \ {\rm and}\ \ \ \ \   \rho_{31} = \frac{\Theta_3}{\Theta_1}\equiv \Theta_{31}.
\end{equation}
The difference of the phase $\Phi_0$ using Eq.(\ref{snr_phase}) for the three detectors will be

\begin{equation}\label{phi-diff}
    \Phi_{21} = \Phi_{0,2} - \Phi_{0,1} \ \ \ {\rm and}\ \ \ \ \   \Phi_{31} = \Phi_{0,3} - \Phi_{0,1}
\end{equation}
The quantities $\rho_{21}, \rho_{31}, \Phi_{21}, \Phi_{31}$ depend only on the position on the sky, and polarization and inclination angle of the binary. Equations (\ref{theta-ratios}) and (\ref{phi-diff}) provide  constraints on these four variables. We assume that the measurement errors on the signal to noise ratios is Gaussian with the standard deviation $\sigma_{\rho}=1$, and the error on the measurement of the phase is $\sigma_{\Phi}=\pi/\rho$. We denote the probability density of the measured SNR ratios as $P(\rho_k)$ and $P(\Phi_k)$.
We calculate the probability density of the SNR ratios as 
$P(\rho_{k1}) = \int d\rho_k d\rho_1 \delta(\rho_{k1}-\rho_k/\rho_1) P(\rho_k) P(\rho_1)$, and the 
phase difference $P(\Phi_{kj})  = \int d\Phi_k d\Phi_j \delta(\Phi_{kj}-(\Phi_k-\Phi_j)) P(\Phi_k) P(\Phi_j)$.

Considering the first set of data to be $ D_1=\rho_{21}, \rho_{31}, \Phi_{21}, \Phi_{31}$. We use the Bayes theorem to obtain the constraints on the position in the sky and on the polarization and inclination angles, where we denote $\Omega_{sky}\equiv (\theta,\phi)$ and $\Omega_{source} \equiv (\iota, \psi)$, 

\begin{eqnarray}\label{fourangles}
    P(&& \Omega_{sky},\Omega_{source}|D_1, I) = \nonumber\\
    && = \frac{P(\Omega_{sky},\Omega_{source}|I) P(D_1| \Omega_{sky},\Omega_{source},I)}{P(D_1|I)}.
\end{eqnarray}

We assume that the prior probability $P(\Omega_{sky},\Omega_{source}|I)$ is uniform on the celestial sphere and on the sphere parameterized by the angles $\iota$ and $\psi$. The likelihood is
\begin{widetext}
\begin{equation}
\begin{split}
P( D_1 | \Omega_{sky},\Omega_{source},I) & = \int d\rho_{21} P(\rho_{21}) 
\int d\rho_{31}P(\rho_{31}) \int d\Phi_{21} P(\Phi_{21})\int d\Phi_{31}P(\Phi_{31}) \delta\left(\rho_{21}-\rho_{21}(\Omega_{sky},\Omega_{source})\right)\\
& \times \delta\left(\rho_{31}-\rho_{31}(\Omega_{sky},\Omega_{source})\right) 
\delta\left(\Phi_{21}-\Phi_{21}(\Omega_{sky},\Omega_{source})\right)
\delta\left(\Phi_{31}-\Phi_{31}(\Omega_{sky},\Omega_{source})\right)
\end{split}
\end{equation}
\end{widetext}

Given the information on sky maps $\Omega_{eff} \equiv (\Omega_{sky},\Omega_{source})$, we have
\begin{equation}\label{thetaeffprob1}
P(\Theta_{eff}|D_1, I) = \int d\Omega_{eff} P(\Theta_{eff}, \Omega_{eff}|D_1, I)
\end{equation}
where,  
\begin{equation}
\begin{split}
    P(\Theta_{eff}, &\Omega_{eff}|D_1, I)  = \\
    & = P(\Omega_{eff}|D_1, I)P(\Theta_{eff}|\Omega_{eff},D_1, I) \\ 
    & = P(\Omega_{eff}|D_1, I)\\
    & \times \left(\frac{P(\Theta_{eff}| I)P(\Omega_{eff},D_1|\Theta_{eff}, I)}{P(\Omega_{eff},D_1| I)}\right)
\end{split}
\end{equation}
Assuming flat prior on $\Theta_{eff}$,
\begin{equation}\label{thetaeffprior}
    P(\Theta_{eff}| I) = \frac{1}{\Delta \Theta_{eff}^{max}}
\end{equation} 
where $\Theta_{eff}^{max} = 6$, we have

\begin{eqnarray}\label{thetaeffprob}
    P(\Theta_{eff}|D_1, I) &\propto& \frac{1}{\Delta \Theta_{eff}^{max}} \int d\Omega_{eff} P(\Omega_{eff}|D_1, I) \nonumber\\
    & & \times \delta\left(\Theta_{eff} - \Theta_{eff}(\Omega_{eff})\right)
\end{eqnarray}

Substituting Eq. (\ref{fourangles}) in (\ref{thetaeffprob}) gives the probability density for $\Theta_{eff}$, where we included explicitly the dependence of $\Theta_{eff}$ on the position in the sky, polarization, and inclination.

We now proceed to impose constraints on the chirp mass and redshift to the source ; i.e., we want to break the degeneracy between the two parameters.  The measured waveform from a detected binary will be analyzed with the best waveform models available when the ET is in operation. For now, we assume conservatively  that these models will be as good as  the models currently available. The analysis of a waveform from a single detector yields at least the redshifted chirp mass, the frequency of the transition from the inspiral to merger and ringdown, and the constraints on spins of the inspiraling bodies. Note that constraints on the spins that depend on the signal to noise are not affected by redshifting the entire signal. The errors on all these quantities will depend on the SNR  of the detected  signal.

We will use the measured value of the red shifted chirp mass ${\cal M}_z$, and we assume that the measured probability density of the this quantity is again Gaussian with the width of $\sigma_{{\cal M}_z}={\cal M}_z /\rho_{eff}$, which we denote as $P({\cal M}_z)$. Although this will be an overestimate in the case of 3G detectors, for fixed SNR, as the larger bandwidth will further decrease the uncertainty because of the much larger number of waveform cycles that will be in band which the templates need to match, the error on the detector frame chirp mass plays a minor role in determining the error on the redshift, distance, and chirp mass of the sources. These errors are mainly affected by the  large  error on $\Theta_{eff}$ . 

\begin{figure}
     \centering
     \includegraphics[width = \columnwidth]{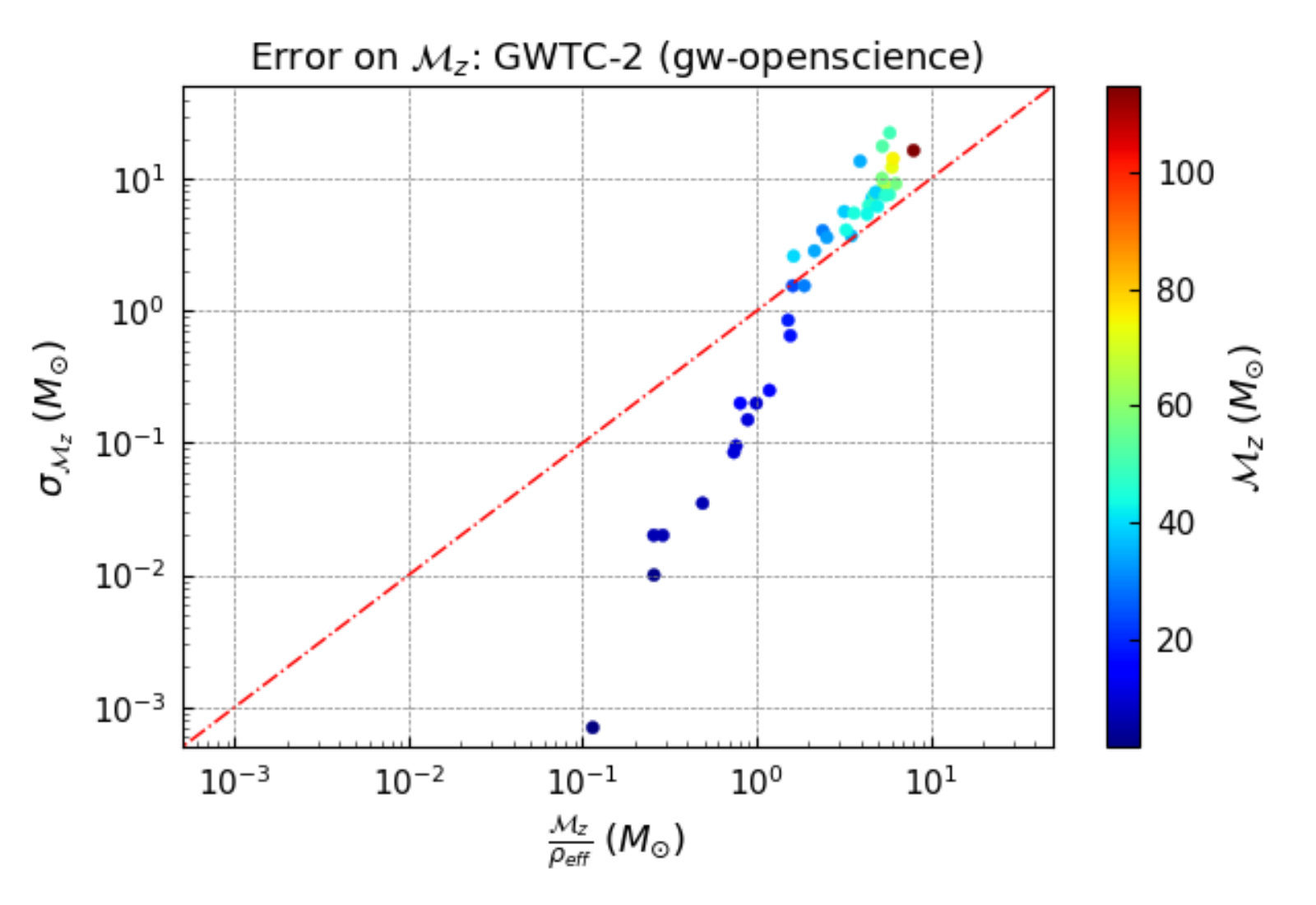}
\caption{The 40 LIGO-Virgo detections from \href{https://www.gw-openscience.org/eventapi/html/GWTC/}{gw-openscience} for which $\mathcal{M}_z$ is mentioned, are plotted here. The plot shows the comparison of the error on $\mathcal{M}_z$ with the $(\mathcal{M}_z/\rho_{eff})$. The color represents the redshifted chirp mass $\mathcal{M}_z$ of the detected compact binary system. The red dashed line represents $\sigma_{{\cal M}_z}={\cal M}_z /\rho_{eff}$. \label{fig:O3a_err}}
\end{figure}

In order to verify our assumption about the magnitude of the error of the redshifted chirp mass $\mathcal{M}_z$, we analyze the errors for 40 LIGO-Virgo detections obtained from \href{https://www.gw-openscience.org/eventapi/html/GWTC/}{"gw-openscience"}, in Fig. \ref{fig:O3a_err}. The plot shows the comparison of $\sigma_{{\cal M}_z}$ with the quantity $\mathcal{M}_z/\rho_{eff}$, and it shows that the assumption $\sigma_{{\cal M}_z}={\cal M}_z /\rho_{eff}$ is a good conservative  approximation for compact binaries in the inspiral phase.

The values of $\sigma_{{\cal M}_z}$ for GWTC-2 detections have been obtained considering the full parameter space, including the spin of the components. It can be seen that although we do not consider the spin values in our analysis, the estimate of the uncertainty of the redshfted chirp mass already includes the spin effects. A quick look at Fig. \ref{fig:O3a_err} shows that the  assumption in our analysis analysis gives an overestimate on the errors on the estimates of the binary parameters, so one can treat the results as conservative.

We assume that the error on $f_{max}$ is small and neglect it. In a complete analysis, one would have access not just to the inspiral but to the full inspiral-merger-ringdown signal, and in that case, $f_{max}$  can be determined rather well. Apart from Eqs. (\ref{fmax}) and (\ref{snreff}), we use the constraint implied by the definition of the chirp mass,
\begin{equation} \label{massratio}
    \frac{{\mathcal{M}}}{M} = \left[\frac{q}{(1+q)^2} \right]^{3/5}< 4^{-3/5}
\end{equation}
where the last inequality comes from setting $q=1$. We denote  this constraint as $I_Q$.\\
Using the Bayesian theorem, 
\begin{equation}\label{qconstraint}
    P({\cal M}, z|I_Q, I) = P({\cal M}|I) P({z}|I) \frac{P(I_Q| {\cal M}, z,I)}{P(I_Q|I)}
\end{equation}
where
\begin{equation}
\begin{split}
    P(I_Q| {\cal M}, z,I) & = \mathcal{H} \left(\frac{\mathcal{M}}{2.8M_{\odot}}\frac{f_{max}(1+z)}{785 \rm {Hz}}\right) \\
    & \times \mathcal{H}\left(4^{-3/5} - \frac{\mathcal{M}}{2.8M_{\odot}}\frac{f_{max}(1+z)}{785 \rm {Hz}}\right)
\end{split}
\end{equation}
and ${\cal H}(x) $ is the Heaviside function.  We can now include the information on the measurement of the redshifted chirp mass, effective SNR, and constraint on the angular factor $\Theta_{eff}$ to impose further constraint on ${\cal M}$ and $z$. Assuming  $D_2 = \mathcal{M}_z, f_{max}, \rho_{eff}$,
\begin{widetext}
\begin{equation}\label{chirpzjointprob1}
    P(\mathcal{M}, z|D_1, D_2, I_Q, I)  = \int d\Theta_{eff}P(\mathcal{M}, z, \Theta_{eff}|D_1, D_2, I_Q, I)
\end{equation}
where,
\begin{eqnarray}
    P(\mathcal{M}, z, \Theta_{eff}| D_1, D_2, I_Q, I) & = & P(\Theta_{eff}| D_1, D_2, I_Q, I)P(\mathcal{M}, z| \Theta_{eff}, D_1, D_2, I_Q, I) \nonumber\\
    & & = P(\Theta_{eff}| D_1, I)\left( \frac{P(\mathcal{M}, z|I_Q, I)P(\Theta_{eff}, D_1, D_2| \mathcal{M},z, I_Q,I)}{P(\Theta_{eff}, D_1, D_2| I_Q,I)} \right)
\end{eqnarray}
where we have used the fact that $D_2$ and $I_Q$ provide no new information on $\Theta_{eff}$,
\begin{equation}\label{chirpzjointprob}
\begin{split}
   P(\mathcal{M}, z|D_1, D_2, I_Q, I) & \propto P(\mathcal{M}, z|I_Q, I) \int \int \int d{\cal M}_z  d\rho_{eff} d\Theta_{eff} P(\Theta_{eff}| D_1,I) P({\cal M}_z)P(\rho_{eff})\\
    & \times \delta({\cal M}_z-{\cal M}(1+z))\delta(\rho_{eff}-\rho_{eff}({\cal M},z,\Theta_{eff}))
\end{split}
\end{equation}

Here, we have used the probability density of $\rho_{eff}$,
\begin{equation}
      P(\rho_{eff})= \int d\rho_1 \int d\rho_2 \int d\rho_3 P(\rho_1)P(\rho_2)P(\rho_3) \delta\left(\rho_{eff}-
      \sqrt{\rho_1^2+\rho_2^2+\rho_3^2}\right),
\end{equation}
\end{widetext}
and expressed Eq. (\ref{snreff}) explicitly as a function of ${\cal M}$, $z$, and $\Theta_{eff}$. Substituting Eqs. (\ref{thetaeffprob}) and (\ref{qconstraint}) in (\ref{chirpzjointprob}) gives the required distribution for $\mathcal{M}, z$.
Taking $D' \equiv (D_1, D_2)$ and $I' \equiv I_Q, I$, the marginalized distribution for $\mathcal{M}$ and  $z$ using Eq. (\ref{chirpzjointprob}) is given as
\begin{equation}\label{zprob}
    P(z|D', I') = \int d \mathcal{M} P(\mathcal{M}, z| D', I')
\end{equation}
and 
\begin{equation}\label{chmprob}
    P(\mathcal{M}|D', I') = \int dz P(\mathcal{M}, z| D', I')
\end{equation}
The distribution for luminosity distance is obtained form the redshift using the cosmological assumptions mentioned in Sec. \ref{cosmoprior},
\begin{equation}\label{dlporb}
    \frac{dP}{dD_L} = \frac{dP}{dz}\frac{dz}{dD_L}
\end{equation}
Finally, we can constrain the mass ratio using Eqs. (\ref{fmax}) and (\ref{massratio}) to express the mass ratio as a function of $\cal M$ and $z$,
\begin{equation}
     q({\cal M},z) = \frac{\xi}{2}-1-\sqrt{\frac{\xi^2}{4}-\xi}
\end{equation}
where 
\begin{equation}
     \xi=\left( \frac{785{\rm Hz}}{f_{max}} \frac{2.8{\rm M}_\odot}{{\cal M}_z} \right)^{5/3}
\end{equation}

\begin{equation}\label{qprob1}
    P(q|D',I') = \int\int d \mathcal{M} dz P(\mathcal{M}, z, q| D', I')
\end{equation}
where
\begin{equation}
\begin{split}
    P(\mathcal{M}, z, q| D', I') & = P(\mathcal{M}, z| D', I') P(q|\mathcal{M}, z, D', I')\\
    & = P(\mathcal{M}, z| D', I') \\
    & \times \left( \frac{P (q|I')P(\mathcal{M}, z, D'|q, I')}{P(\mathcal{M}, z, D'| I')}  \right)
\end{split}
\end{equation}
We assume a flat prior on the mass ratio $P(q|I') = 1$, so
\begin{equation}\label{qprob}
    P(q|D',I') \propto \int \int d\mathcal{M} dz P(\mathcal{M}, z| D', I') \delta(q - q(\mathcal{M}, z))
\end{equation}
Substituting Eq. (\ref{chirpzjointprob}) in (\ref{qprob}) gives the probability density for mass ratio $q$. Thus, using the observed signal to noise ratios, phases, redshifted chirp mass, and $f_{max}$, we have constrained the astrophysical parameters such as  the chirp mass, redshift, and mass ratio of the binary. Neglecting the error of $f_{max}$ only influences the estimate of the mass ratio and the total mass of the binary. However, these errors are dominated by the errors of the redshift for stellar mass binaries. Including this error requires adding one more integration over the probability density of $f_{max}$.

\begin{figure*}
\includegraphics[width = \columnwidth]{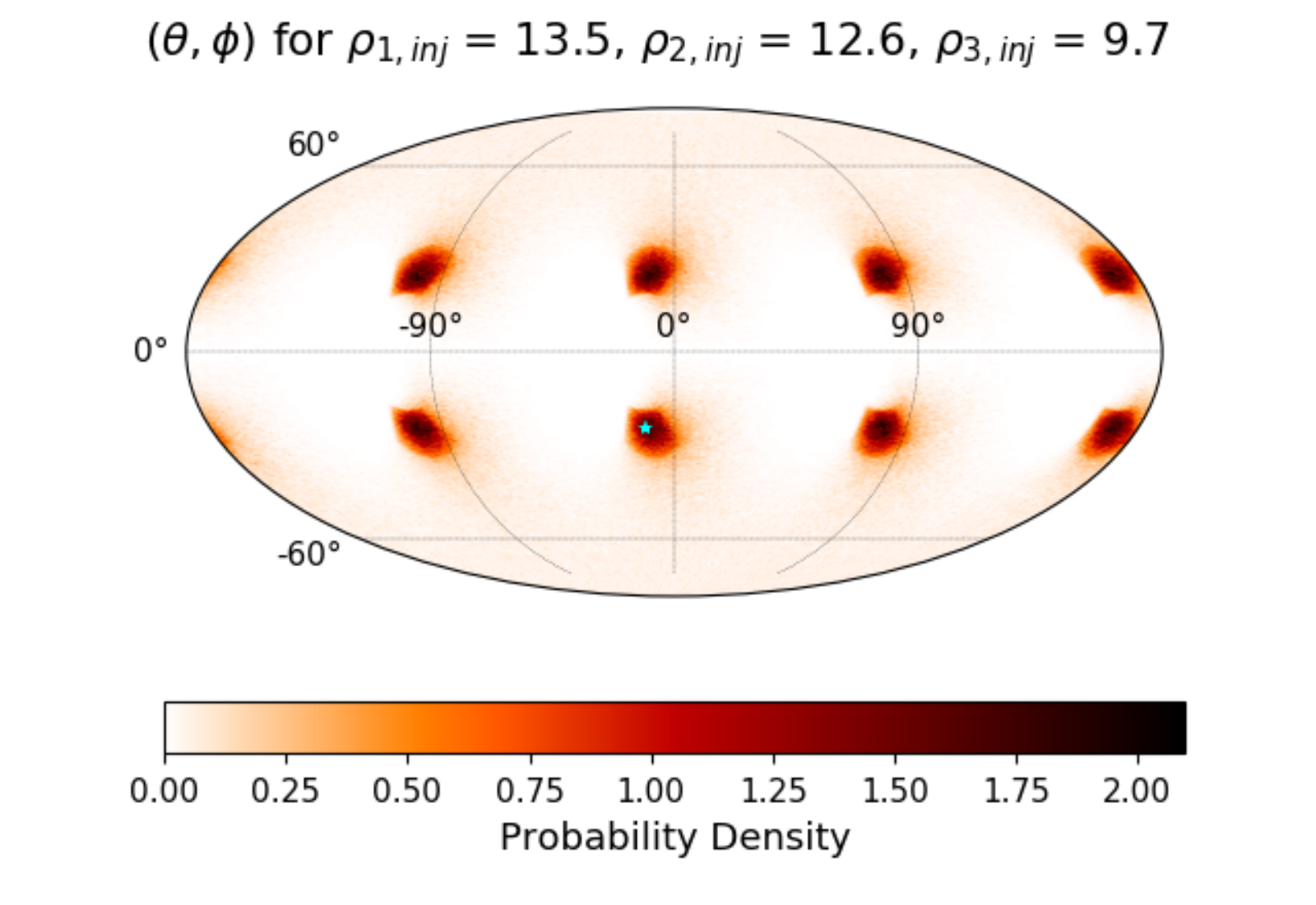}
\includegraphics[width = \columnwidth]{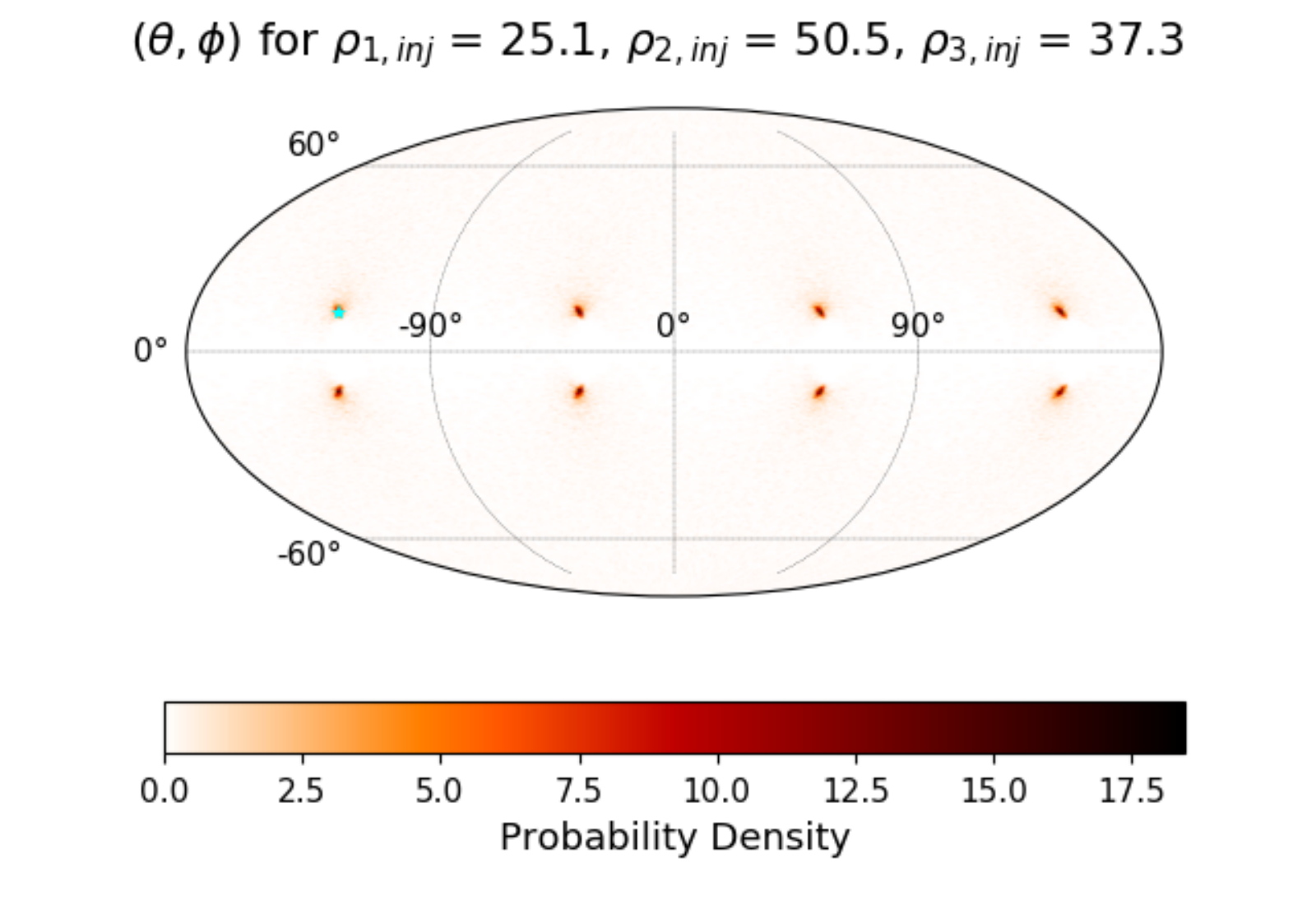}\\
\includegraphics[width = \columnwidth]{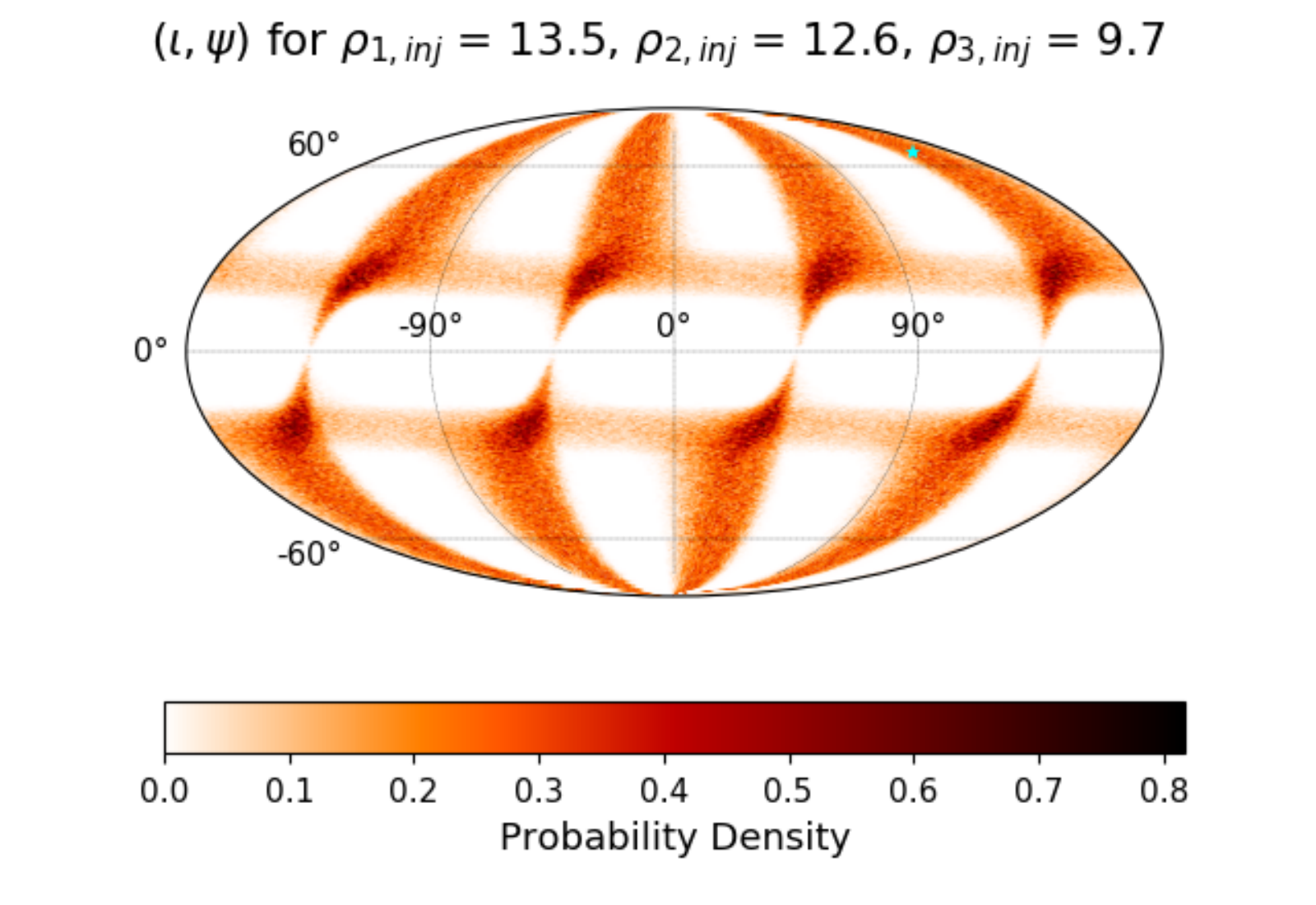}
\includegraphics[width = \columnwidth]{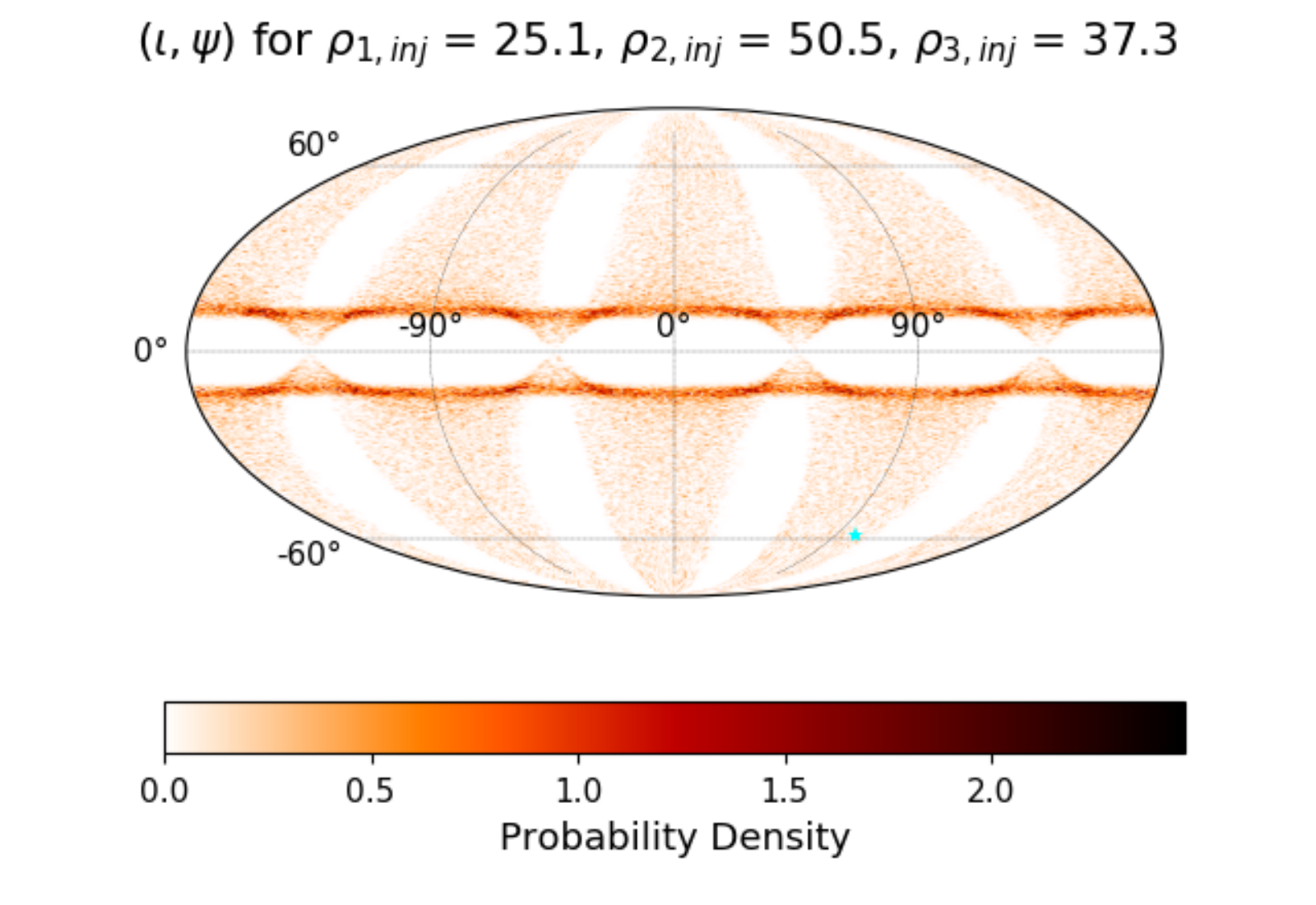}
\caption{ Localization for $(\theta, \phi)$ and $(\iota, \psi)$ recovered for $\rho_{eff} = 20.89$(left) and  for $\rho_{eff} = 67.64$(right). The blue star denotes the coordinates for the injected source.  For case 1 (left), the area for $90 \%$ probability about the peak for $(\theta, \phi)$ is spread over 13398.91 sq degs, whereas for case 2 (right), the spread is reduced to 9818.80 sq degs. The values of the parameters for both these cases are mentioned in Table \ref{tab:casedetail}. \label{fig:localization}}
\end{figure*}

\begin{figure*}
\includegraphics[width = \columnwidth]{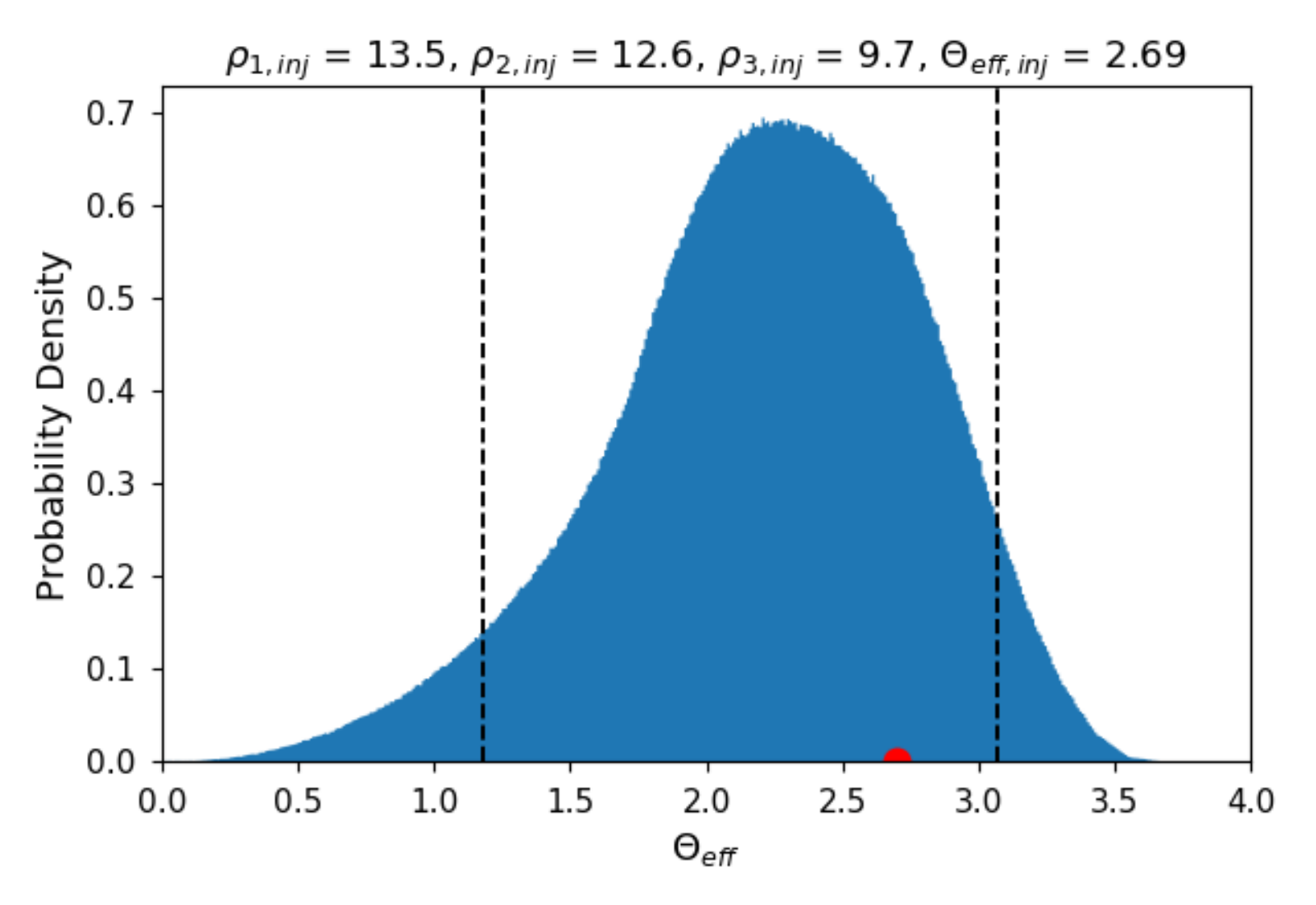}
\includegraphics[width = \columnwidth]{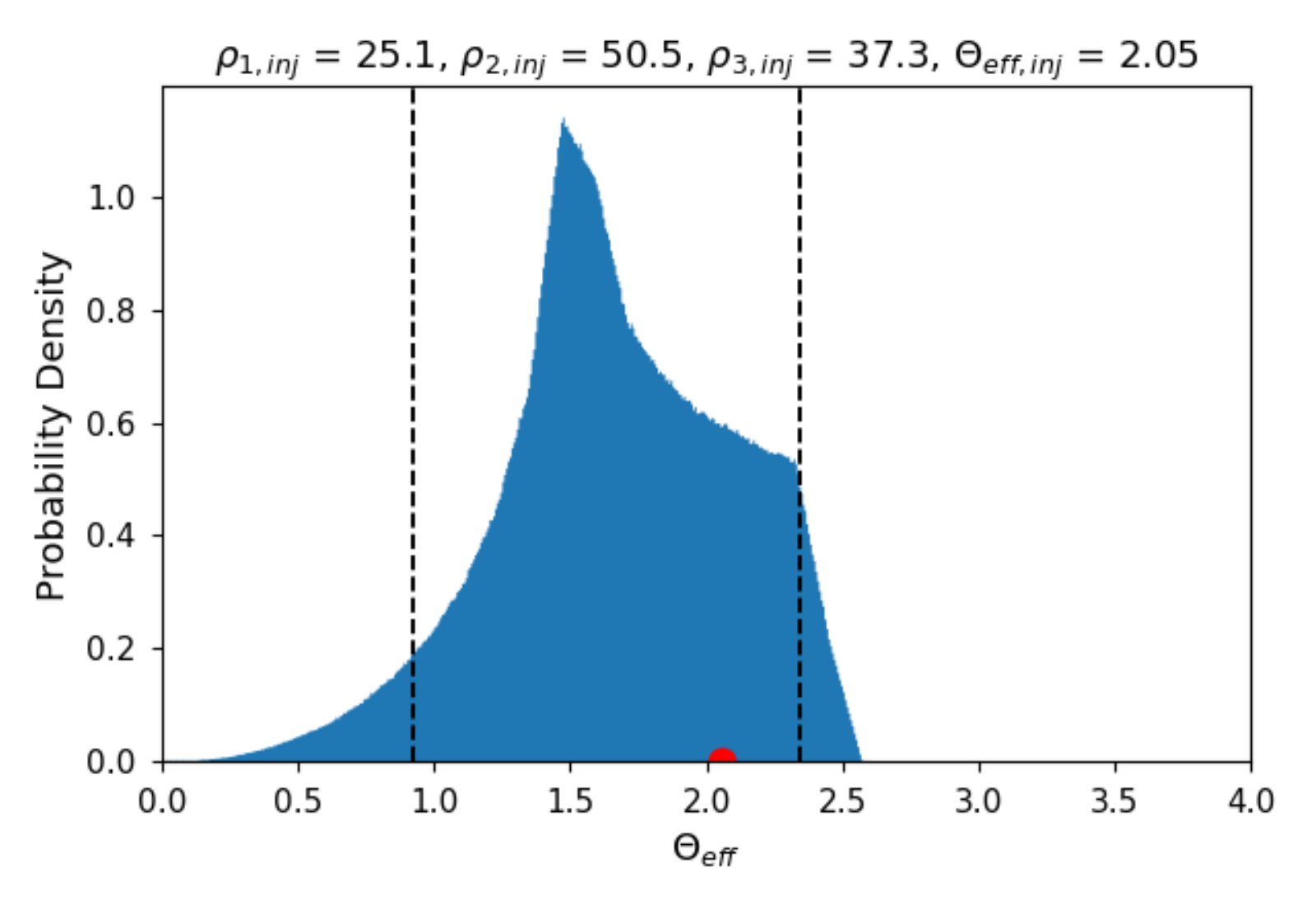}
\caption{Plots showing the probability for $\Theta_{eff}$ recovered for  case 1 (left) and  for case 2 (right). The red dot denotes the value of the parameter of the injected source in each plot. The black dashed vertical lines show the limits of $90\%$ probability region about the median. \label{fig:probdist0}}
\end{figure*}

\section{Mock Source Catalog}\label{sec:injections}

We create a mock binary black hole population assuming that the distributions of masses, distances, locations in the sky, and polarizations are independent.  In particular, we assume that the mass distribution is the same for all distances.

\subsection{Mass distribution}\label{massprior}
We generate binary black hole sources using the one-parameter power law to model the distribution of primary component masses \cite{PhysRevX.6.041015},
\begin{equation}
p(m_1|\alpha) = m_1^{-\alpha} ;\;\alpha = 2.35
\end{equation}
The mass ratio $q = m_2/m_1$ is chosen uniformly for the range [0,1]. Both the component masses $m_1$ and $m_2$ are constrained such that $m_1, m_2 > 5 \; M_{\odot}$. The total mass is also restricted $M \leq 100 \;M_{\odot}$.

\subsection{Angular distribution}
The response function for the detector and the quantity $\Theta$ depend on the four angles $\theta, \phi, \iota, \psi$ as seen from Eqs. (\ref{antenna}) and (\ref{theta}). We choose $\cos\theta, \phi/ \pi, \cos \iota$, and $\psi/ \pi$ to be uncorrelated and distributed uniformly
over the range $[-1,1]$. The quantities $\Theta_i$ for each of the three ET detectors are then calculated using Eq. (\ref{theta}). The values of $\Phi_0$ for the three detectors is obtained using equation (\ref{snr_phase}).

\subsection{Cosmological assumptions}\label{cosmoprior}
In this work, we assume flat cosmology with $\Omega_m=0.3$,  $\Omega_m + \Omega_\Lambda = 1, $ $\Omega_k = 0$, $H_0 = 67.3 \; \rm km s^{-1} \;\rm Mpc^{-1}$ \cite{2015PhRvD..92l3516A}. The relation between the luminosity distance $D_L$ and redshift $z$ is obtained using the analytic approximation given by \cite{Adachi2012}. For this cosmology, the variation of comoving volume $V$ with redshift $z$ is given as, 

\begin{subequations}
\begin{equation}
  \frac{dV}{dz} = 4\pi D_H\frac{D_L^2}{(1+z)^2E_z},
\end{equation}
where,
\begin{equation}
D_H = c/H_0 \;  {\rm and} \\ \; E_z = \sqrt{\Omega_m(1+z)^3 + (1-\Omega_m)}
\end{equation}
\end{subequations}
We generate BBH sources spread in redshift in the range $[0.001, 10.0]$ assuming  that the rate density of mergers per unit comoving volume per unit time  is constant in the Universe. This is an unphysical assumption; however, we wish to use this to demonstrate the capabilities of ET in a model unbiased way. The realistic population will include evolution of the merger rate density with redshift.

\section{Results}\label{sec:results}

\begin{figure*}
\includegraphics[width = \columnwidth]{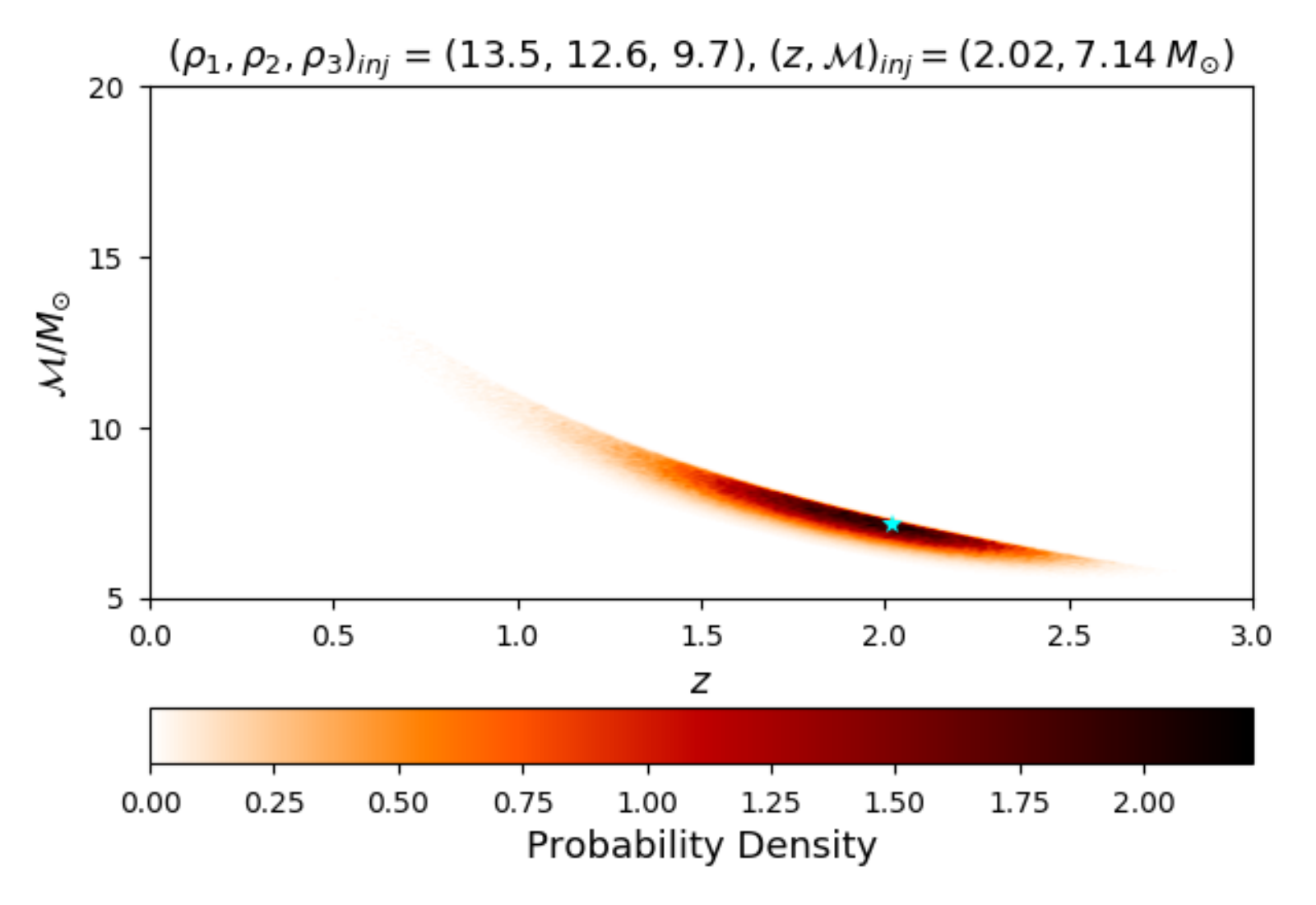}
\includegraphics[width = \columnwidth]{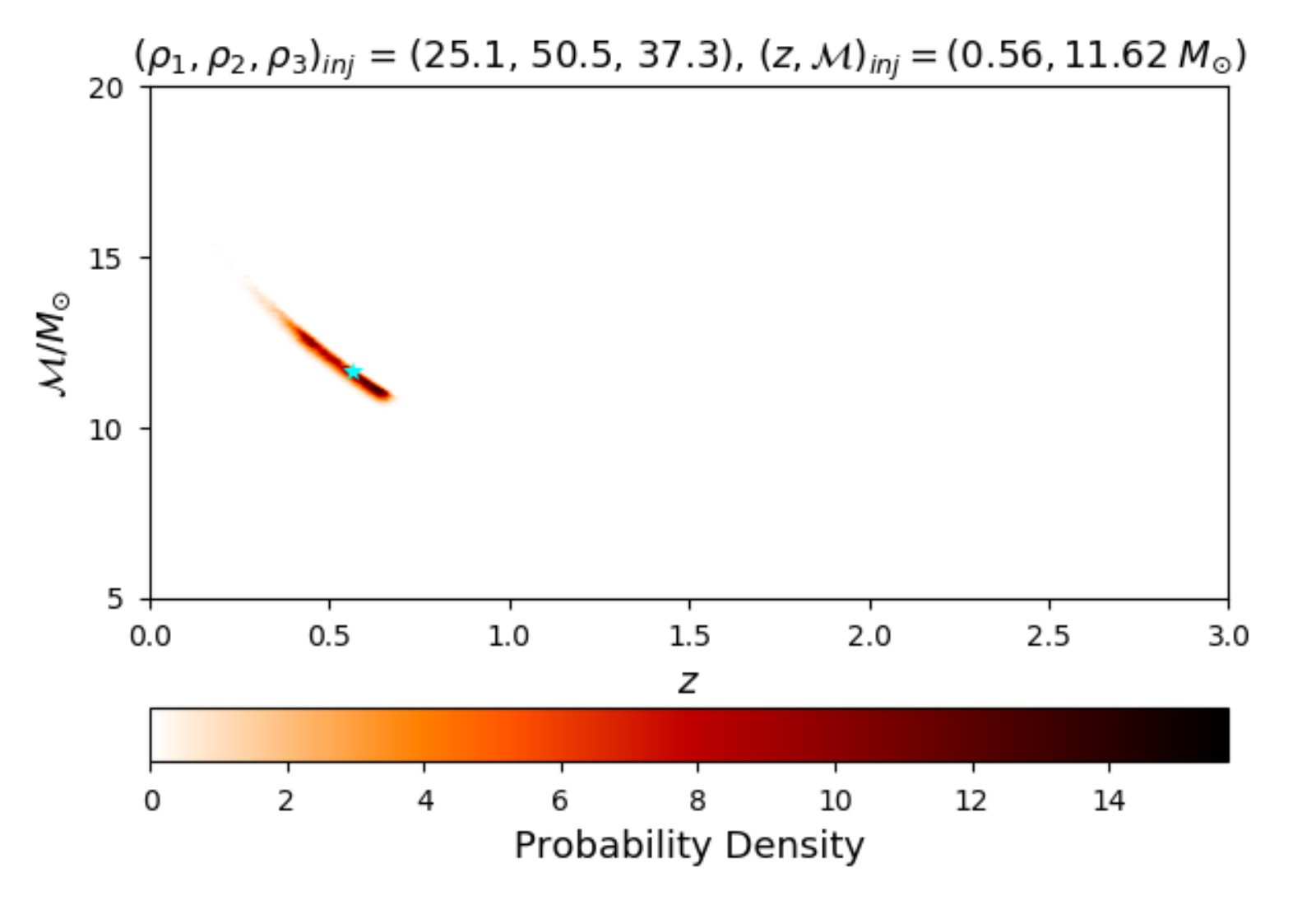}
\caption{The joint probability for chirp mass $\mathcal{M}$ and redshift $z$, recovered for case 1 (left) and case 2 (right). The blue star denotes the values for the injected source. The values of the parameters for both these cases are mentioned in Table \ref{tab:casedetail}. \label{fig:z_chm}}
\end{figure*}

\begin{figure*}[ht!]
\includegraphics[width = \columnwidth]{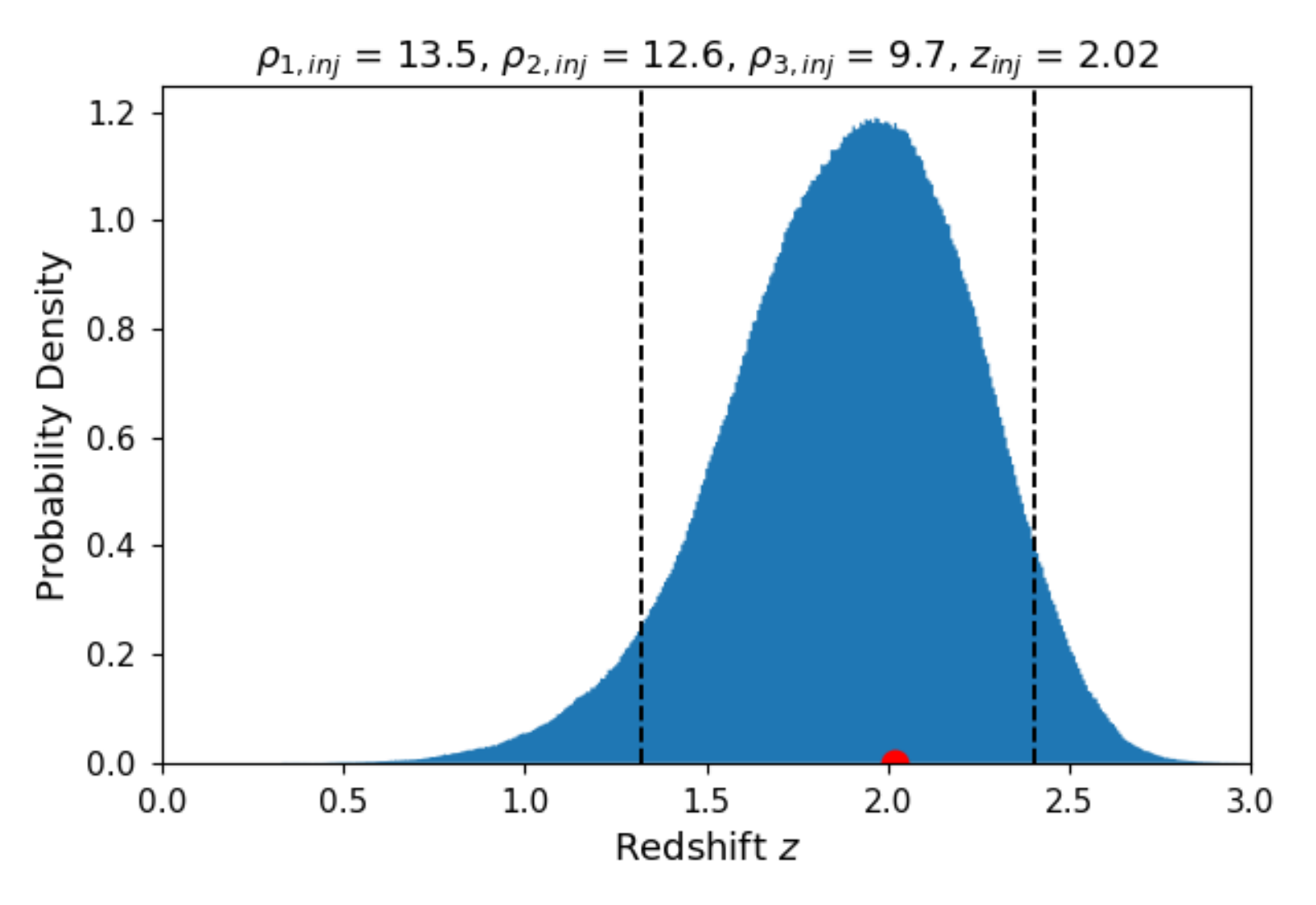}
\includegraphics[width = \columnwidth]{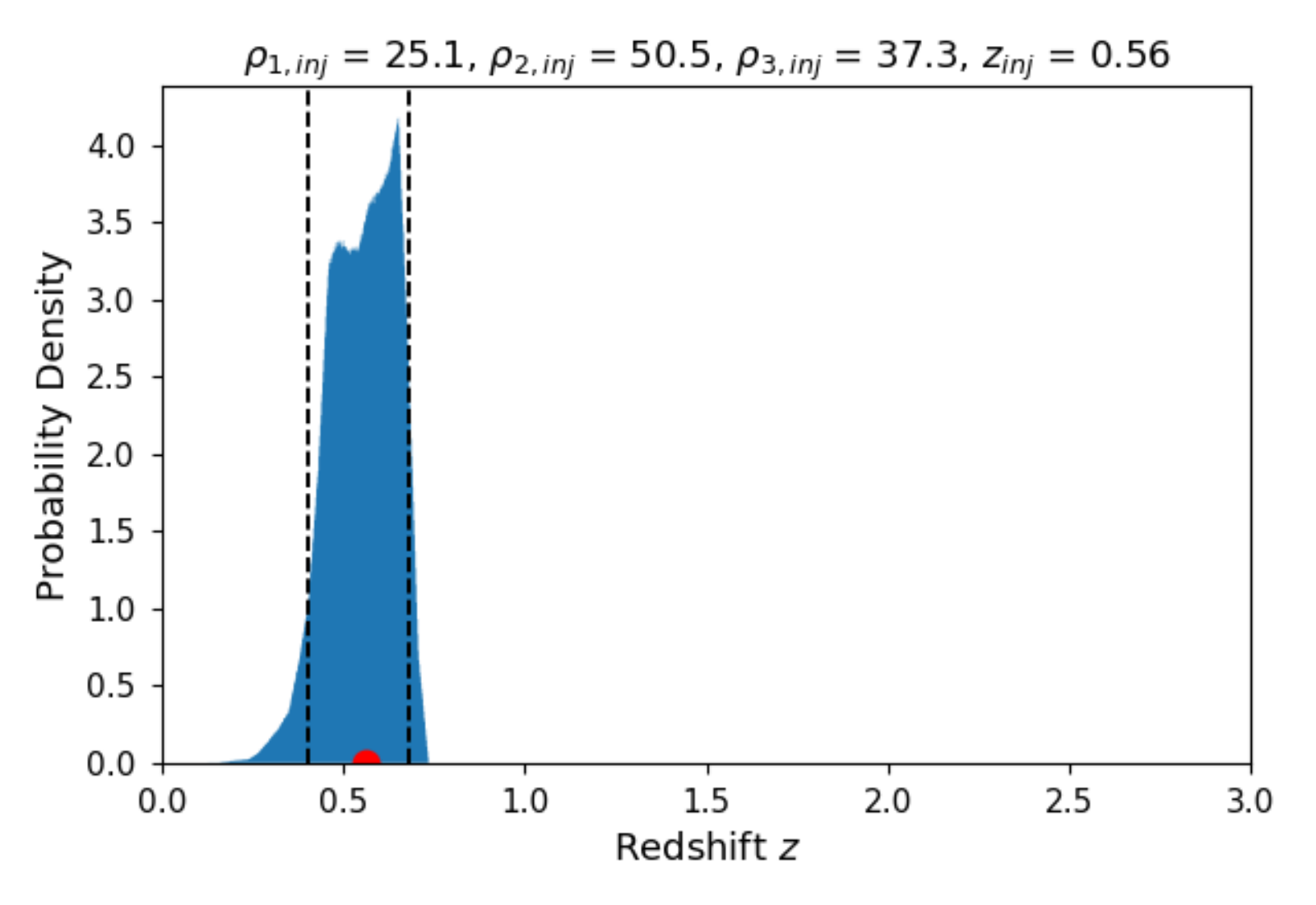}\\
\includegraphics[width = \columnwidth]{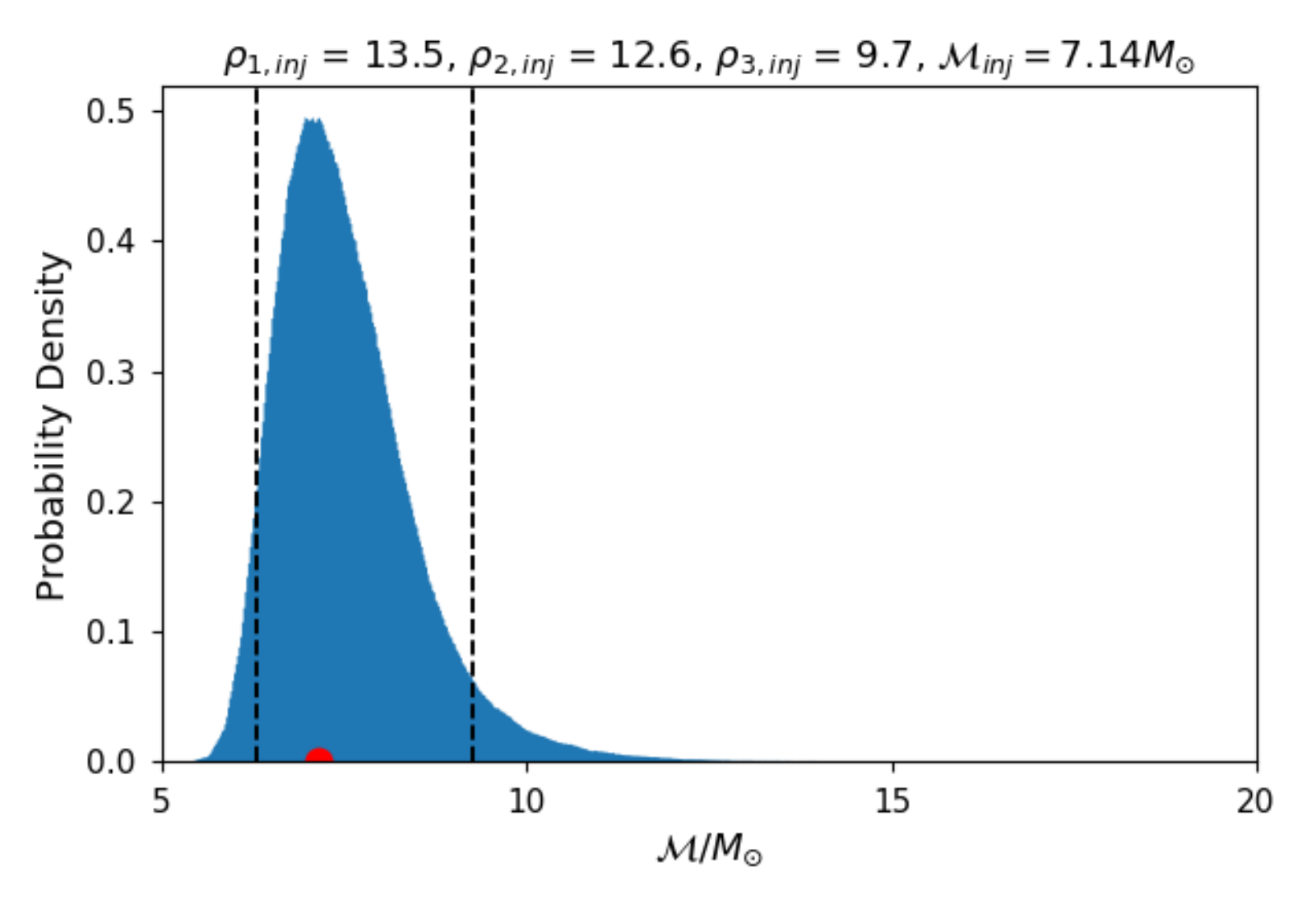}
\includegraphics[width = \columnwidth]{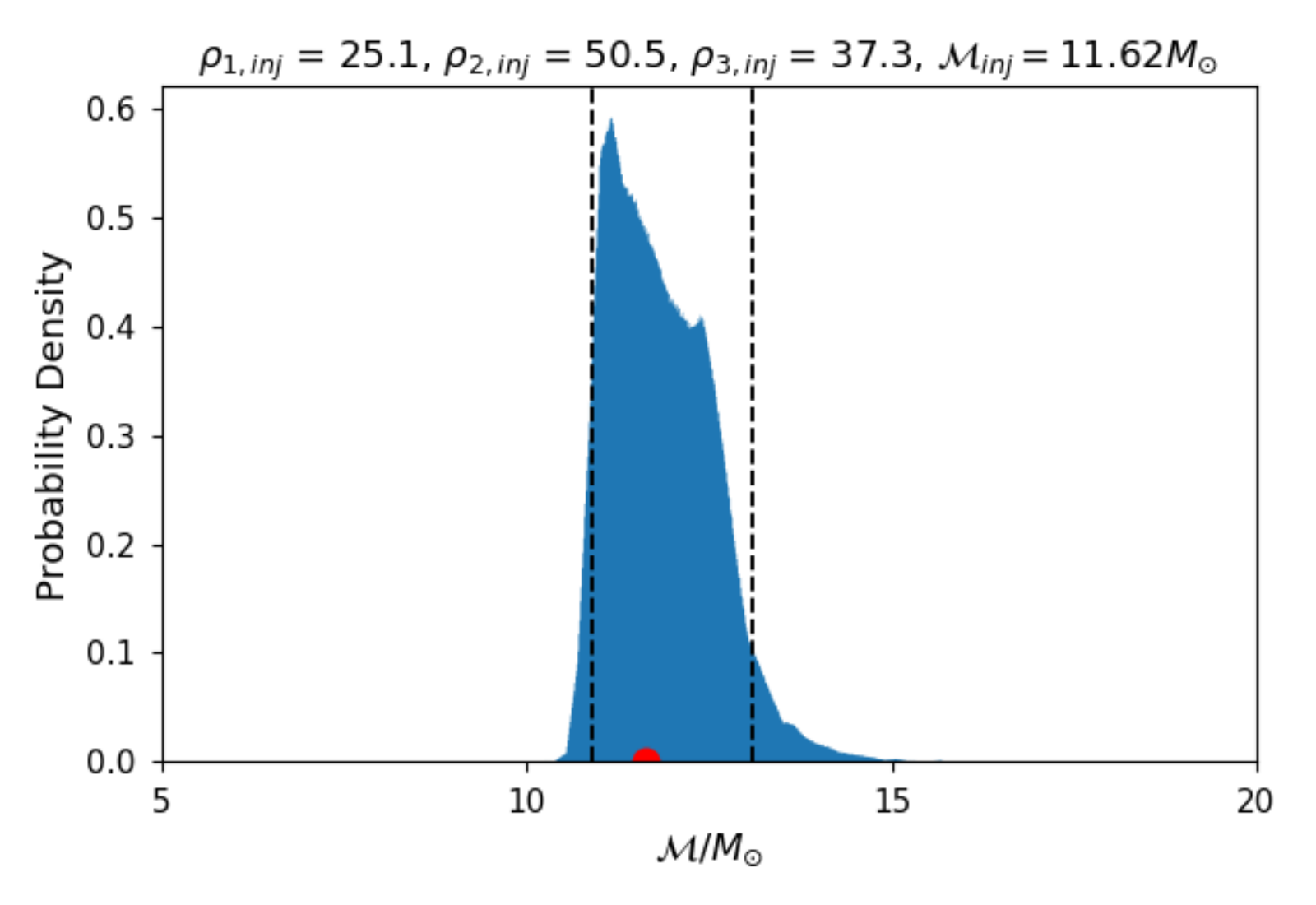}\\
\caption{Plots showing the marginalized probability for redshift $z$  and  $\mathcal{M}$ recovered for  case 1 (left) and  for case 2 (right). The red dot denotes the value of the parameter of the injected source in each plot. The black dashed vertical lines show the limits of $90\%$ probability region about the median. \label{fig:probdist1}}
\end{figure*}

\begin{figure*}
\includegraphics[width = \columnwidth]{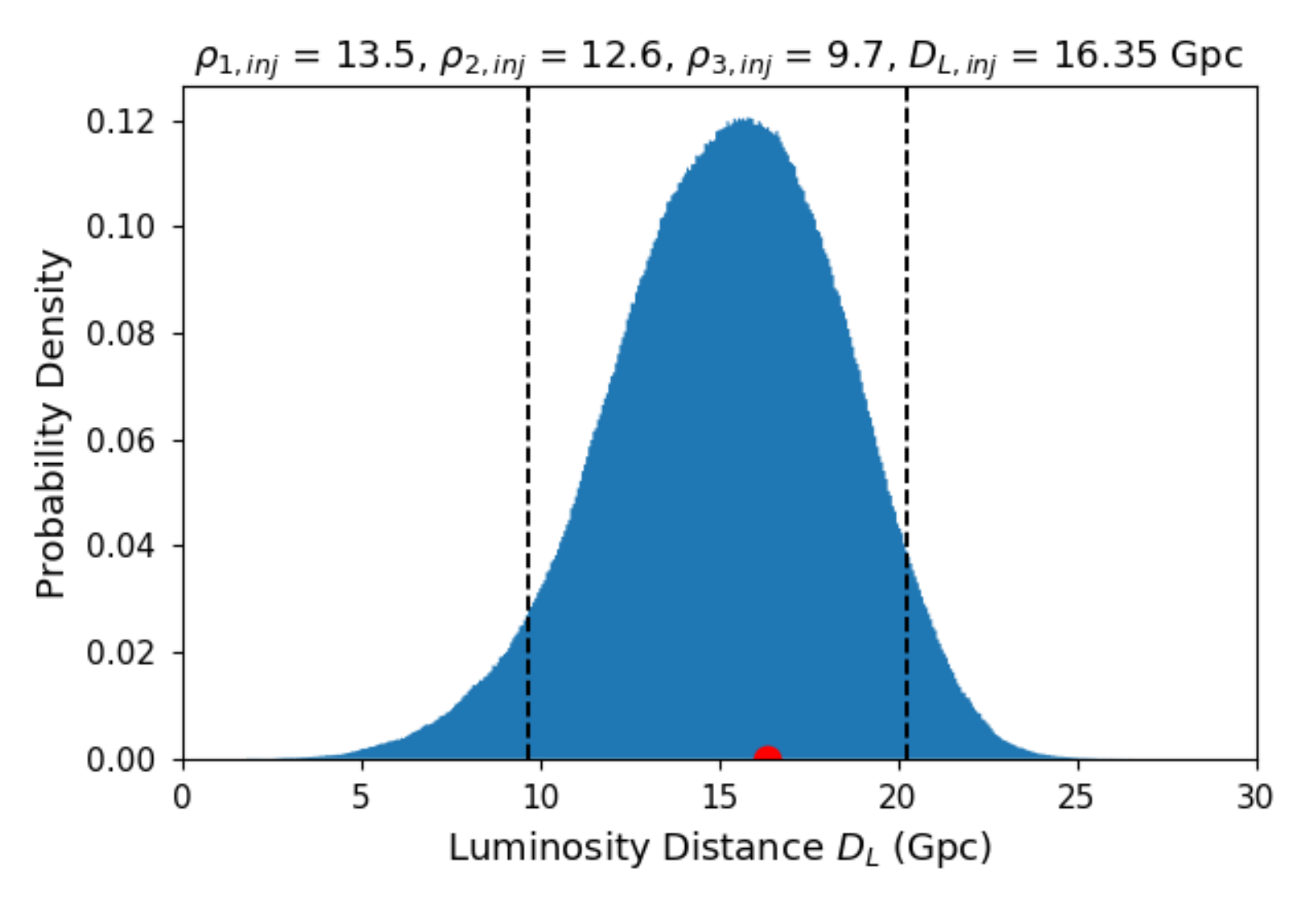}
\includegraphics[width = \columnwidth]{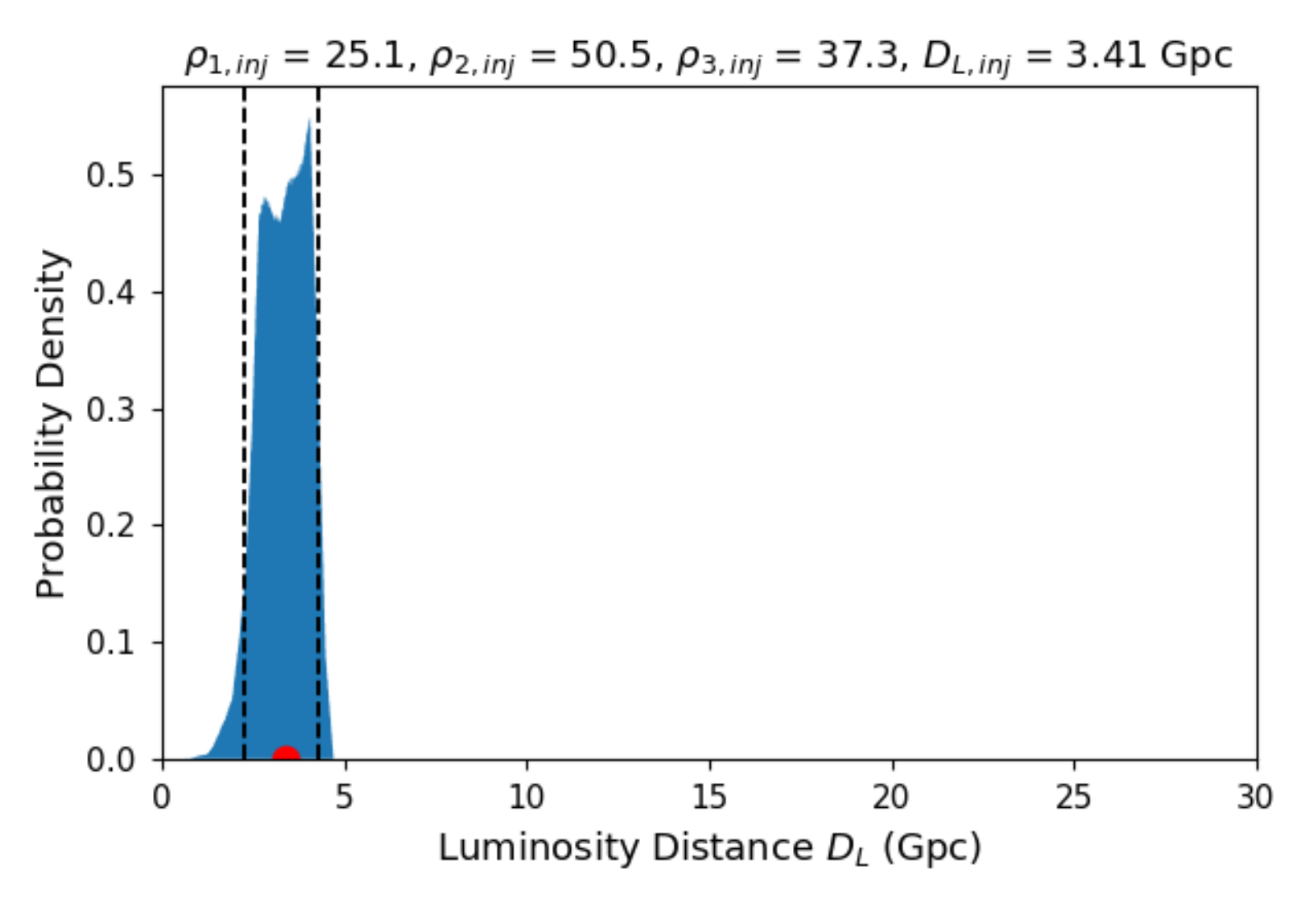}\\
\includegraphics[width = \columnwidth]{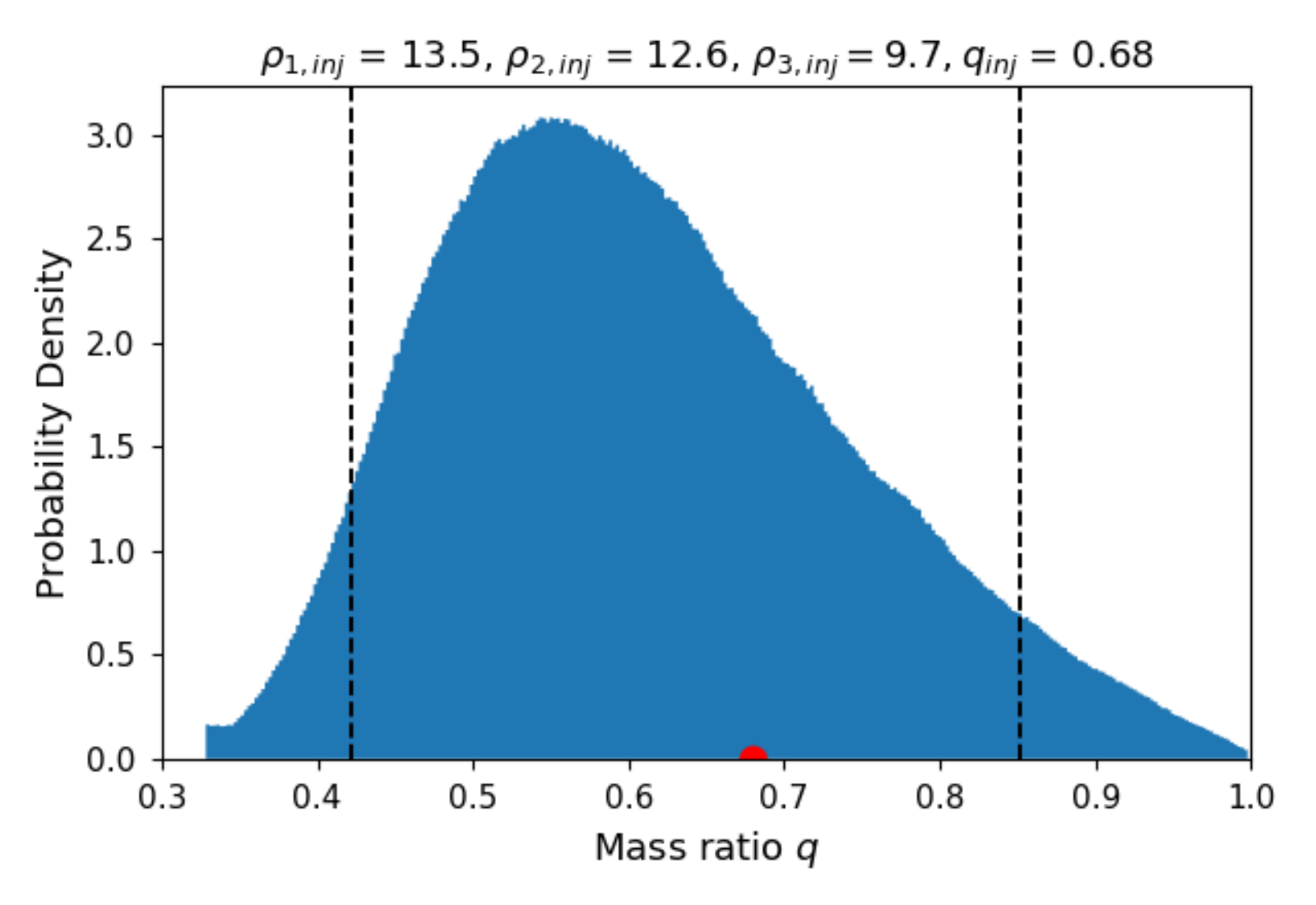}
\includegraphics[width = \columnwidth]{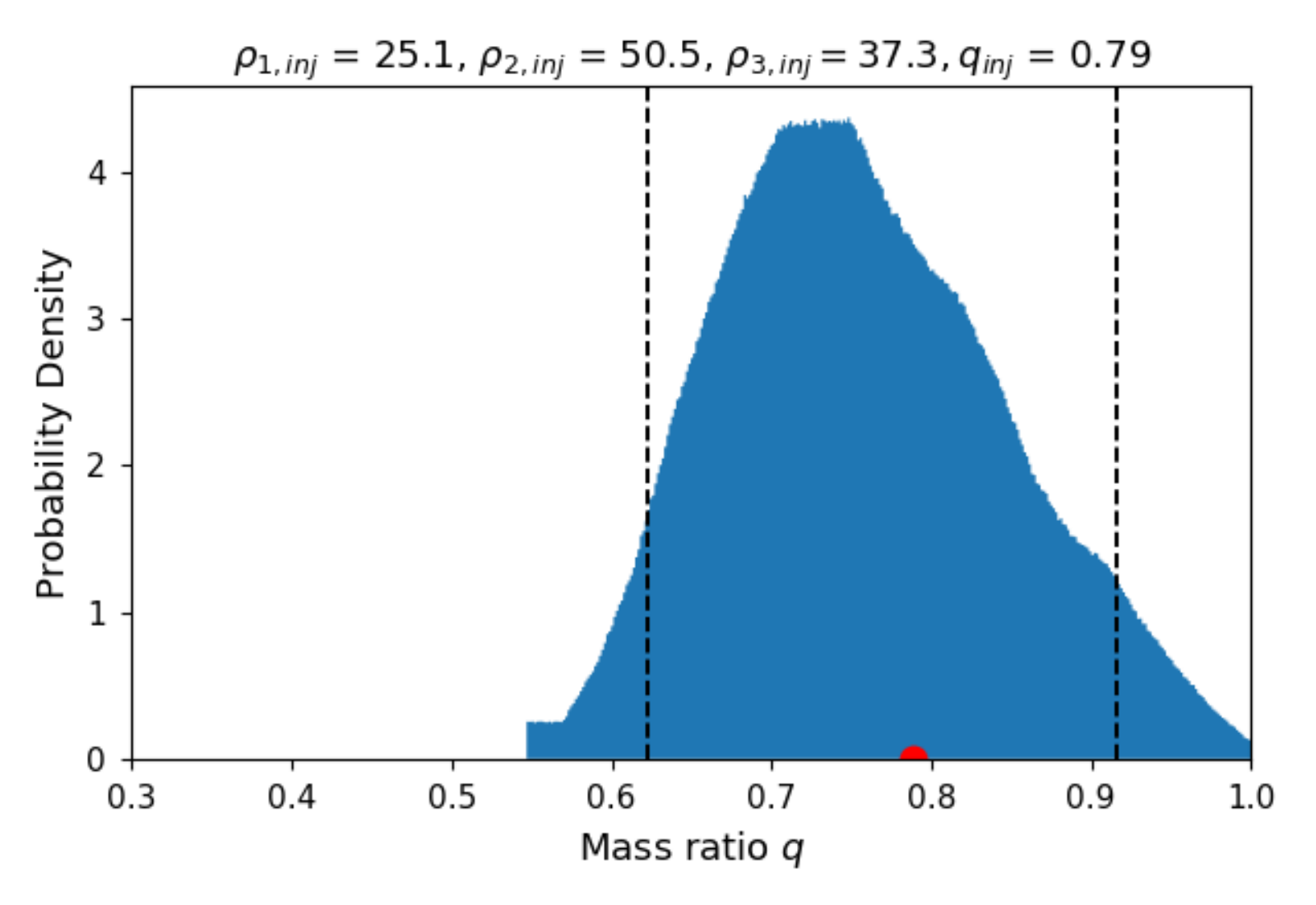}\\
\caption{Plots showing the probability for $D_L$  and  $q$ recovered for case 1 (left) and  for case 2 (right). The red dot denotes the value of the parameter of the injected source in each plot. The black dashed vertical lines show the limits of $90\%$ probability region about the median.\label{fig:probdist2}}
\end{figure*}

\begin{table*}
\caption{Details of two injected cases of $\rho_{eff}$ discussed in subsequent figures. \label{tab:casedetail}}

\begin{ruledtabular}
\begin{tabular}{ccccccccccc}
 & $\rho_1$ & $\rho_2$ & $\rho_3$ & $z$ & $D_L$ & $\mathcal{M}$  & $M$ & $q$ & $(\cos\theta, \phi, \cos\iota, \psi)$ \\
 &  &  &  &  & (Gpc) & ($M_{\odot}$) & ($M_{\odot}$) & & \\
\hline
Case 1 & 13.55 & 12.61 & 9.68 & 2.02 & 16.35 & 7.14 & 16.76 & 0.68 & (0.39, -0.19, -0.91,  2.67) \\
Case 2 & 25.06 & 50.53 & 37.34 & 0.56 & 3.41 & 11.62 & 26.91 & 0.79 & (-0.21, -2.19,  0.86, 1.76)  \\
\end{tabular}
\end{ruledtabular}
\end{table*}

\begin{table}
\caption{Errors on the recovered values for the two injected cases mentioned in Table \ref{tab:casedetail}. The error on the parameters is the estimate for the spread of $90 \%$ probability about the median for the respective parameters. The area estimated for the localization of $(\theta, \phi)$  is the spread for $90 \%$ probability about the peak value of the $(\theta, \phi)$ distribution. \label{tab:errordetail}}
\begin{ruledtabular}
\begin{tabular}{ccccccc}
& $\Delta \Theta_{eff}$ & $\Delta \mathcal{M} $ & $\Delta z$ & $\Delta D_L$ &
$\Delta q$ & Area for $(\theta, \phi)$ \\
 & & ($M_{\odot}$) &  & (Gpc) &  & (sq deg)\\ 
\hline
Case 1 & 1.87 & 3.03 & 1.08 & 10.52 & 0.43 & 13398.91 \\
Case 2 & 1.42 & 2.18 & 0.27 & 1.95 & 0.29 & 9818.80 \\
\end{tabular}
\end{ruledtabular}
\end{table}

\begin{figure*}
\includegraphics[width = \columnwidth]{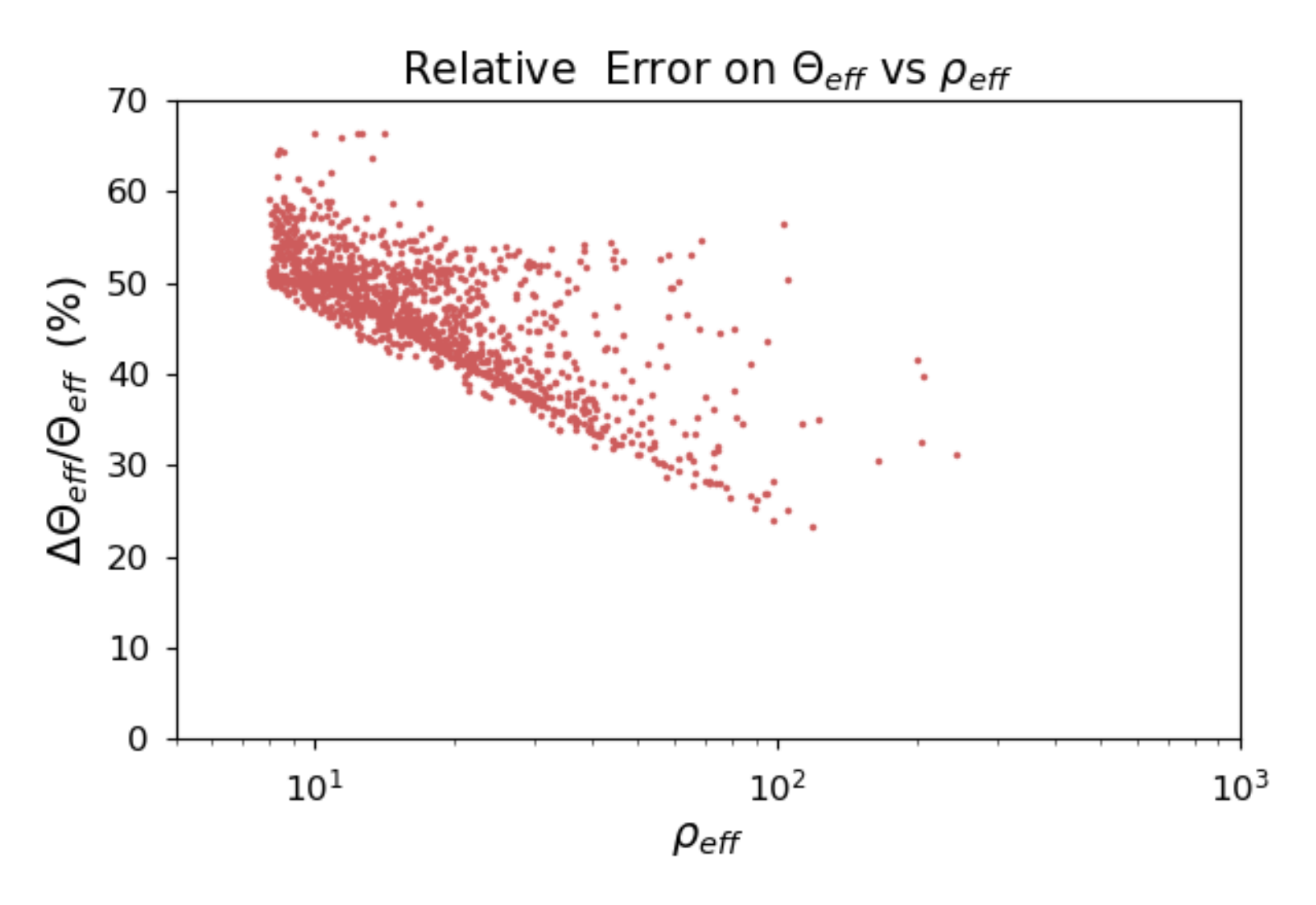}
\includegraphics[width = \columnwidth]{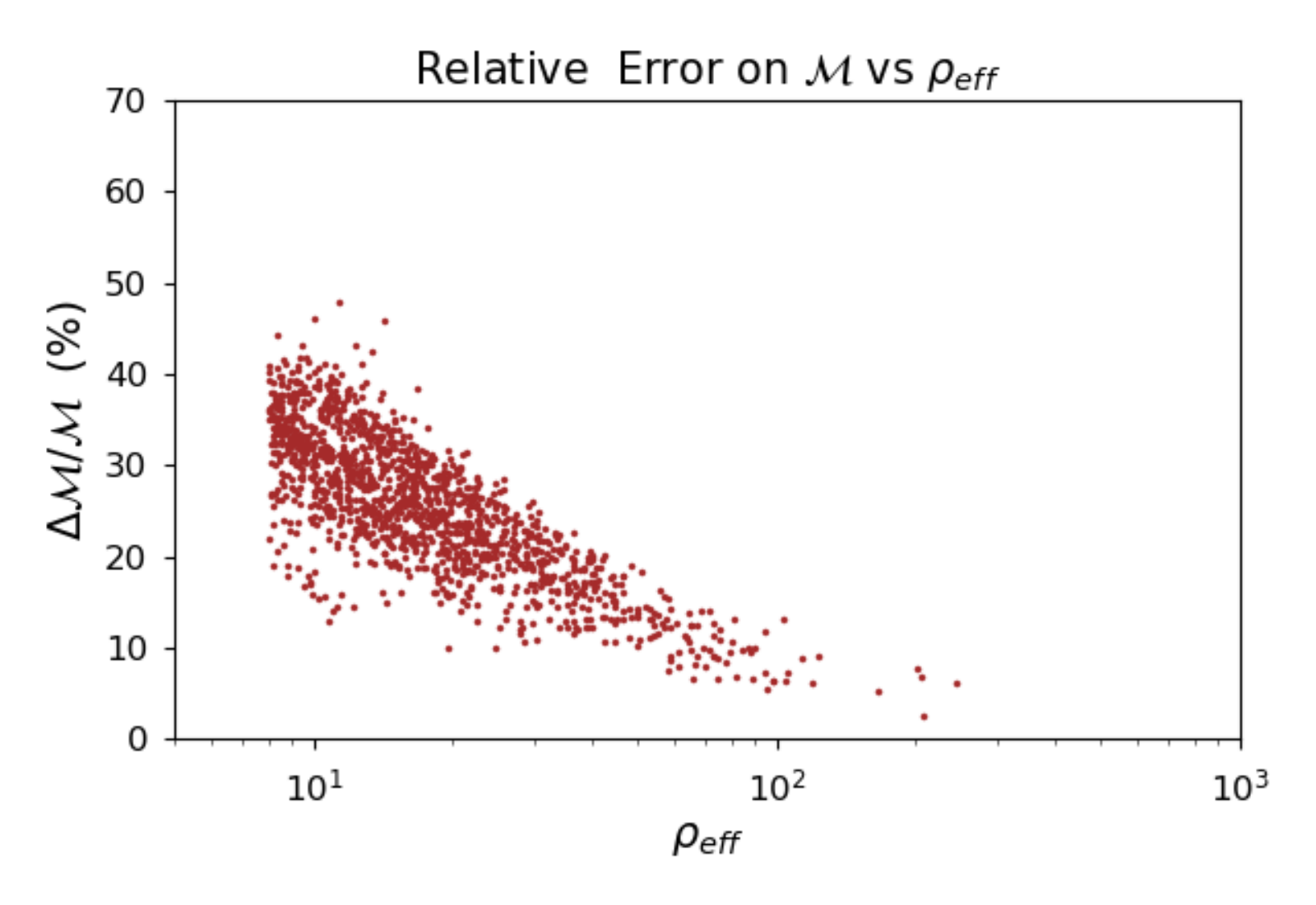}\\

\includegraphics[width = \columnwidth]{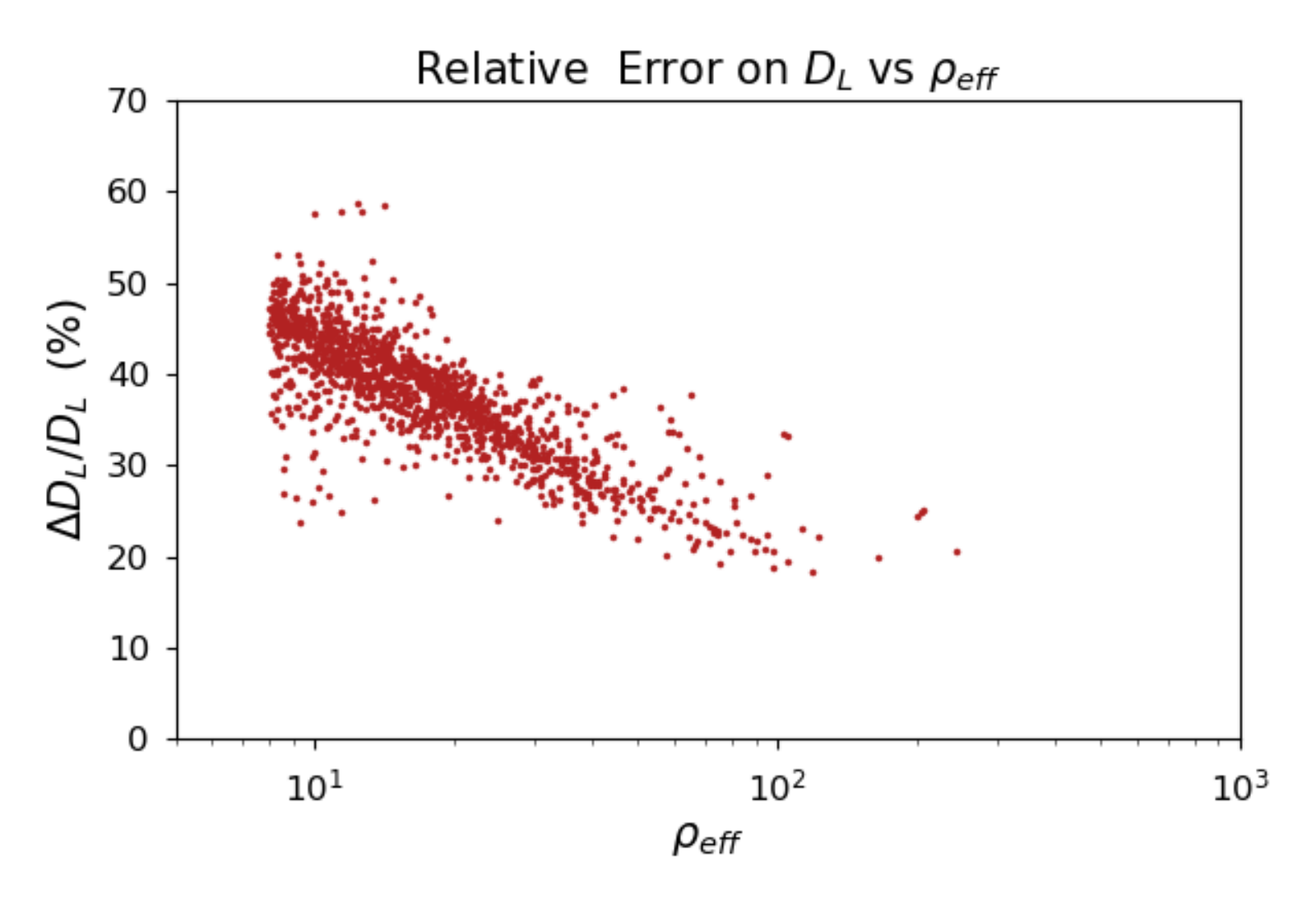}
\includegraphics[width = \columnwidth]{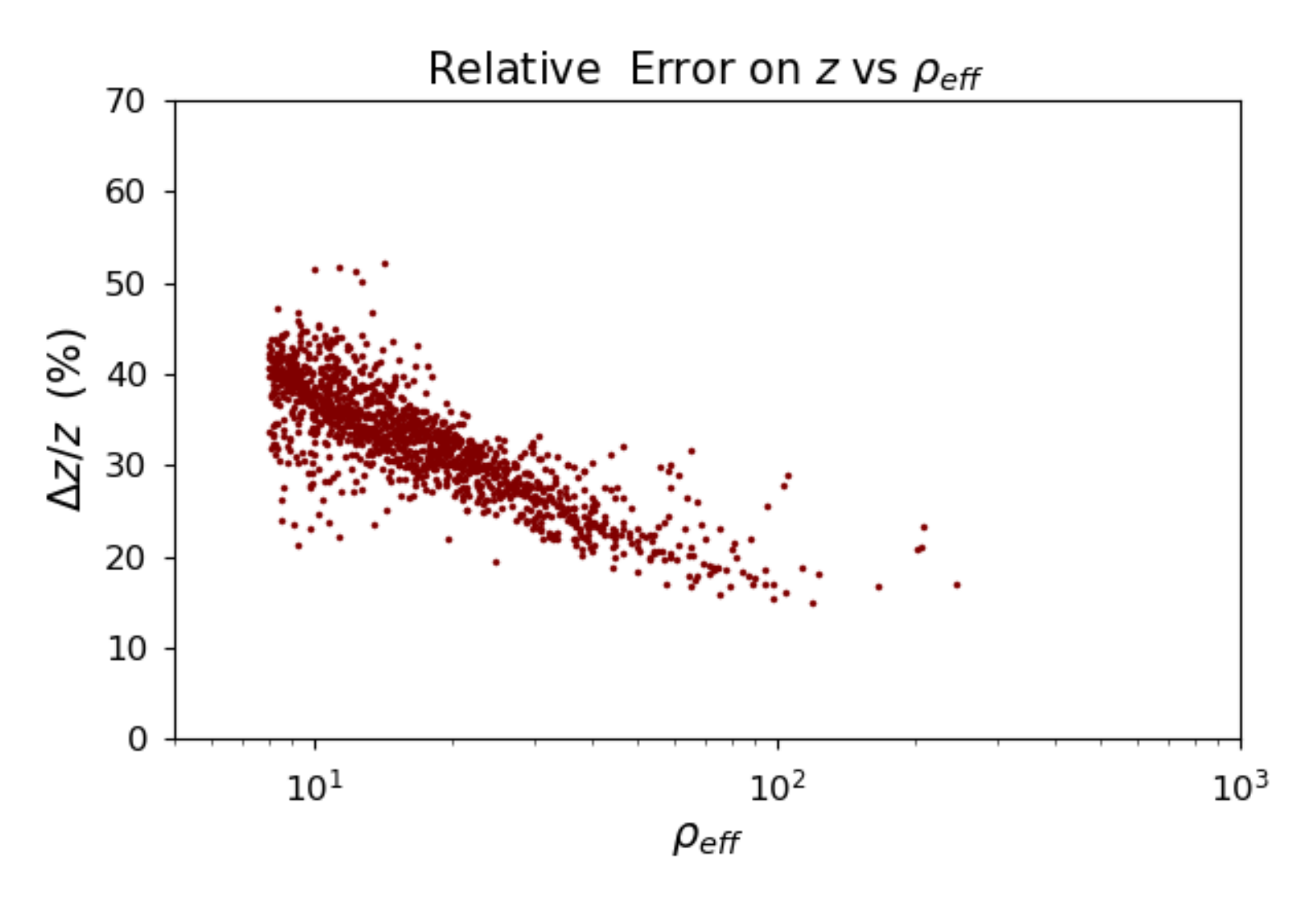}\\

\includegraphics[width = \columnwidth]{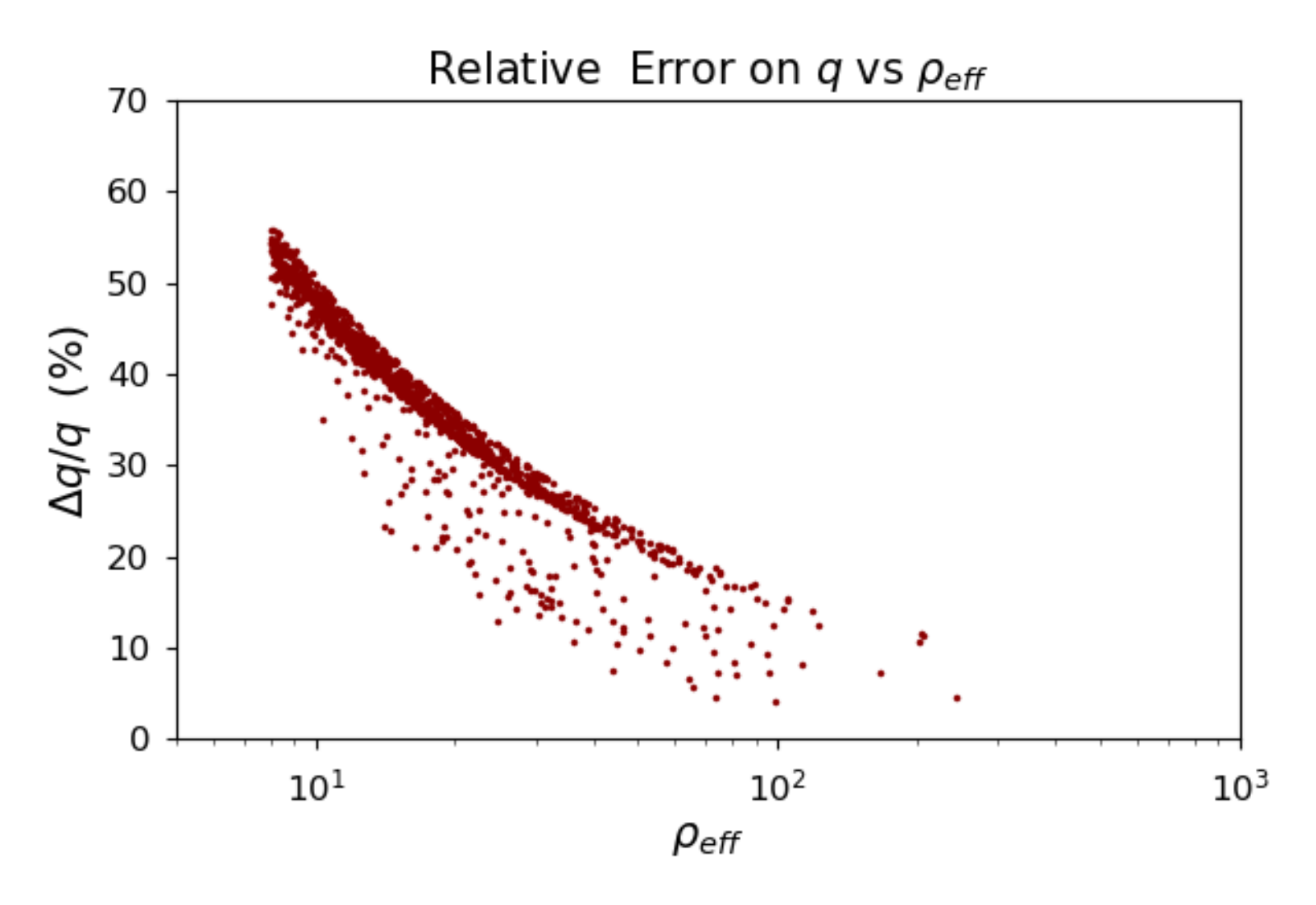}
\includegraphics[width = \columnwidth]{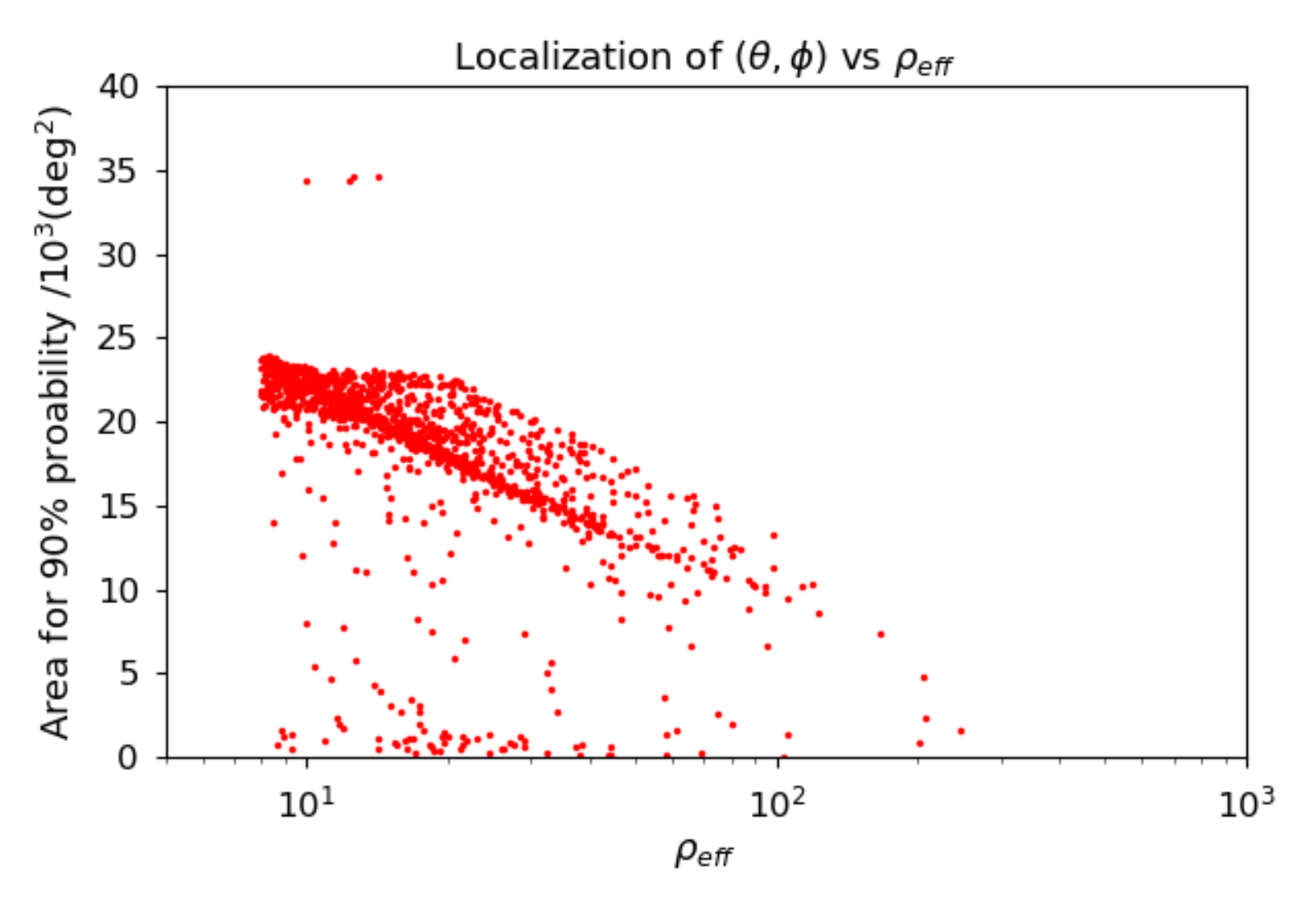}\\

\caption{Relative errors on $\Theta_{eff}, \mathcal{M}, D_L, z, q$ for all the injected sources with $\rho_{eff} > 8$. The relative error on the parameters as shown in (a,b,c,d,e) are estimated from the spread of $90 \%$ probability about the median for the respective parameters. The area estimated for the localization of $(\theta, \phi)$  is the spread for $90 \%$ probability about the peak value of the $(\theta, \phi)$ distribution. \label{fig:errorplots}}
\end{figure*}

The simulation was carried out for 2061 sources, out of which 1359 sources lying above the threshold $\rho_{eff} >8 $ were taken up for analysis. We start by generating a four-dimensional space of $\cos\theta, \phi/ \pi, \cos \iota$ and $\psi/ \pi$, randomly distributed uniformly over the range $[-1,1]$. This 4D space is then constrained using Eqs. (\ref{theta-ratios}) and (\ref{phi-diff}) to give the posterior distribution of $\Theta_{eff}$ specified by Eq. (\ref{thetaeffprob1}). We construct a 2D grid for redshift $z$ and chirp mass $\mathcal{M}$, with ranges from $0.0005 \leq z \leq 15.0$ and $ 4 \leq \mathcal{M}/M_{\odot} \leq 60$.  The assumed recovered redshift range is larger that the one we used for injecting sources, see Sec. \ref{cosmoprior}. This is because we want to allow the errors of the redshift to spread into this larger range so that our error estimates are more realistic.

The prior on $z$ is as explained in Sec. \ref{cosmoprior}. The prior on $\mathcal{M}$ is obtained by fitting an exponential function to that mentioned in Sec. \ref{massprior} such that $ p(\mathcal{M}) \propto 10^{(-\mathcal{M}/17)}/35$. This is done so as to extend both the lower and upper mass cutoff values. This 2D grid of $z$ and $\mathcal{M}$ is limited by the condition given by Eq. (\ref{qconstraint}) and by measured value of $\mathcal{M}_z$. The 3D grid comprising of $z$, $\mathcal{M}$, $\Theta_{eff}$ gives the values of $\rho_{eff}$, which further constrains the $z, \mathcal{M}$ grid within the error spread of the known $\rho_{eff}$ value.

We start by discussing two injections (case 1 and case 2) with the parameters given in Table \ref{tab:casedetail}. We have chosen two binaries with fairly standard chirp masses at typical luminosity distances for the ET. The sky coordinates are given in the detector system connected with interferometer 1.
The first one called case 1 is the case with relatively low signal to noise and the ratios between the SNRs in each detector close to unity. The second one called case 2 is a case with a large effective SNR and the ratios of the SNRs in the detectors far from unity.  
For case 1 and case 2, we recovered the localization in the four parameter angular space (sky and polarization) using Eq. (\ref{fourangles}).  We present the resulting probability densities of $(\theta, \phi)$  and  $(\iota, \psi)$ in Fig. \ref{fig:localization}. We see degeneracy in the recovered angles coming from the nature of dependence of $\Theta$ function on the angles. In particular, there is a symmetry around the equatorial plane ($\theta \rightarrow -\theta$) and also with respect to rotation by $90^{\circ}$ in longitude ($\phi \rightarrow \phi + 90^o $). This results in eight images of possible location of the source as seen in Fig. \ref{fig:localization}. The actual location of the source recovered is to be on one of the images. The orientation and polarization plots show similar structure of eightfold degeneracy. 
We calculate the area spread for $90 \%$ probability about the peak value of the $(\theta, \phi)$ distribution. For case 1, with $\rho_{eff} = 20.89$, the area covering 90\% of probability is spread over 13398.91 sq degrees, whereas for case 2 with $\rho_{eff} = 67.64$, this  is reduced to 9818.80 sq degrees.

The localization in case 2 seems to be much better that in case 1. This is due to SNR ratios being far from unity; however, in this case, the 90\% region is still quite large as there is a significant amount of probability density dispersed around the dark dots in Fig. \ref{fig:localization}. 
The recovered distribution of $\Theta_{eff}$ for case 1 and 2 is shown in Fig. \ref{fig:probdist0}. The fact that we use the information from the three detectors narrows down these distributions and allows to place constraints on the chirp mass and the redshift.

We obtain the joint distribution of  $z, \mathcal{M}$  using Eq. (\ref{chirpzjointprob}) and present the resulting joint probability distribution for both case 1 and case 2  in Fig. \ref{fig:z_chm}. The elongated stripes correspond to the constant redshifted chirp mass for each case, while the constraints coming from the measured effective signal to noise ratio and $\Theta_{eff}$ introduce a weaker constraint along this stripe. The marginalized probabilities for $z$ and  $\mathcal{M}$ are shown in Fig. \ref{fig:probdist1}, obtained using Eqs. (\ref{zprob}) and (\ref{chmprob}). The shape of these distributions is primarily determined by the probability distribution of $\Theta_{eff}$.
We calculate the probability distribution of the luminosity distance using Eq. (\ref{dlporb}) and that of mass ratio using Eq. (\ref{qprob}).
The recovered distribution for the luminosity distance $D_L$ and the mass ratio $q$ is shown in Fig. \ref{fig:probdist2}. The error on the parameters $\mathcal{M}, z, D_L, q$ gives the spread for the $90 \%$ probability region about the median of the respective distributions. The error values for both case 1 and case 2 are listed in Table \ref{tab:errordetail}. 

As mentioned in the beginning of this section the analysis of a population of 2061 sources  yielded  1359 detections. We have recovered their parameters of these binaries using the method outlined above. 
The relative errors on the distribution of $\Theta_{eff}, \mathcal{M}, D_L, z, q$ and the area for localization of $(\theta, \phi)$,  obtained for all the injected sources with $\rho_{eff} > 8$ are shown in Fig. \ref{fig:errorplots}. The relative error on the parameters are estimated from the spread of $90 \%$ probability about the median for the respective parameters. The area estimated for the localization of $(\theta, \phi)$  is the spread for $90 \%$ probability about the peak value of the $(\theta, \phi)$ distribution. It can  be seen in the figures \ref{fig:errorplots} that the relative errors reduce for higher $\rho_{eff}$ and so does the area for localization. The relative errors for $\Theta_{eff}, \mathcal{M}, D_L, z, q$  are $\sim 55 \%, \sim 30 \%, \sim 45 \%, \sim 35 \%, \sim 50 \%$   respectively for $\rho_{eff} \sim 8$ and these are reduced to $\sim 40 \%, \sim 10 \%, \sim 25 \%, \sim 20 \%, \sim 10 \% $ respectively for $\rho_{eff} \sim 100$. The area of localization reduces from $\sim 20\times 10^3$ sq degs to $ \sim 5 \times 10^3$ sq degs as we go from $\rho_{eff} \sim 8$ to 100. 

\section{Conclusion}\label{sec:conc}

We have investigated the capability of using the Einstein Telescope as a single instrument to investigate properties of binary black holes mergers observed by this instrument alone. The observation by ET alone means that one cannot use time of flight delays to constrain the position in the sky of a given source. However, we show that the use of different antenna patterns for each of the three detectors that make ET provides useful information to constrain position in the sky and polarization. What is more important is that one can use the information to constrain the combined detector response function $\Theta_{eff}$. This information in turn, allows us to constrain the redshift, and chirp mass of a merger. In this paper, we have made an assumption that we restrict the analysis to the mergers of short duration in comparison to the Earth rotation. Thus, we assume that the position of the source does not change in the detector coordinates during the observation of a merger.

We have analyzed a mock catalog of stellar mass BBH to study the efficiency of the single ET to constrain the parameters of such binary systems using the ET-D design sensitivity for the analysis. The angles describing the location of the source, its inclination, and polarisation have been constrained using the ratios of the SNRs generated in each of the three detectors of the equilateral triangle configuration of ET. We then used this constraint on the angles to estimate $\Theta_{eff}, \mathcal{M}, D_L, z, q$. In the best case, we found the single ET in triangular configuration can constrain the localization area for $90 \; \%$ probability  region of $(\theta, \phi)$ to a minimum value of 40.86 sq degs, for $\rho_{eff}  = 103.33$. The values of $\mathcal{M}$ and $q$ can also be constrained within $10 \%$ for $\rho_{eff} \sim 100$; however, the typical accuracies are of the order of 25\%.

We have used a simplified model where we take into account only the inspiral part of the waveform. Of course a detailed analysis should take into account all of the phases of coalescence : inspiral, merger, and ringdown. However, we show that the main constraint on the value of $\Theta_{eff}$ comes from the ratios of signal to noise values in the individual detectors of ET or the ratios of the amplitudes of the signals in these detectors. Adding the full waveform will mean that each detection has a slightly higher signal to noise ratio for typical stellar mass binaries. The constraints on the redshift and chirp mass depend on the estimate of $\Theta_{eff}$, but when the signals are strong, these constraints do not improve a lot. Including the merger and ringdown will affect mostly the parameter estimation of the most massive binaries  for which most of the SNR comes in these phases. Therefore, taking in to account the full waveform will not affect the errors estimates for most binaries in this study and will improve the capabilities of the ET to detect and characterize the intermediate mass binaries with masses above $100 M_\odot$.

The results presented above should be put in context by comparing them with the expected uncertainties with detections by networks of interferometers. This has been studied in detail by \citet{PhysRevD.95.064052}. In this paper, the authors have shown that the sky localization of binaries with two third generation detectors should localize binaries with the accuracy better than 100 sq degs, and that the typical error regions should be a few sq degs for binaries at redshifts below 3. We find that the single ET cannot match this accuracy, and the typical error regions span a large portion of the sky with the size of $10^4$ sq degs. However, the distance estimates uncertainties using the network of detectors range from 20\% at redshift below 3 to nearly 80\%  at redshifts of 6 to 20. We show that the single ET can estimate the distance with a similar accuracy, see Fig. \ref{fig:errorplots}. In the case of the estimate of the source frame chirp mass, the single ET estimates are also comparable to the network results. We show that the source frame chirp mass can be estimated with accuracy ranging from 10\% for strong sources to 40\%  near detection threshold, see Fig. \ref{fig:errorplots}, while \citet{PhysRevD.95.064052} estimate the median errors for network in the similar range.   

We conclude that the Einstein Telescope used as a single instrument can be used to investigate properties of merging binaries and allows us to constrain  their distance and chirp mass mass ratios with 20\%-30\% accuracy. The constraints on the position in the sky are not very strong; however, there is a small fraction of binaries for which the position can be constrained to less than 1000 sq degs. While for individual sources these accuracies are quite large, one must state that investigating the properties of the population of compact object binaries will be possible with the Einstein Telescope as a single instrument. The ET will be detecting in excess of 10000 binary black holes per year. Thus, if there was a feature in the compact object mass spectrum, we should be able to improve the determination of the location of such spectrum with an accuracy $\approx \sigma_M  N^{-1/2}$, where $\sigma_M$ is the typical accuracy of mass determination and $N$ is the number of detections. Assuming a typical mass relative accuracy of 25\%, we should be able to determine details of the mass spectrum within 0.25\%. A detailed study of the capabilities of ET for determination of mass spectra of compact object binaries is in preparation. 

\begin{acknowledgments}
We acknowledge the support from the Foundation for Polish Science grant TEAM/2016-3/19 and NCN grant UMO-2017/26/M/ST9/00978.
We thank Chris Van Den Broeck for useful comments and suggestions. This document has been assigned Virgo document number VIR-0876B-20.
\end{acknowledgments}

\newcommand{\apjl}{Astrophysical J. Lett.}
\newcommand{\aap}{Astronomy and Astrophysics}

\bibliography{ET_BBH}
\bibliographystyle{apsrev}

\end{document}